\documentclass[12pt,a4wide]{article}
\pdfoutput=1
\usepackage{jheppub}
\usepackage{amsfonts}
\usepackage{amscd}
\usepackage{amssymb}
\usepackage{amsmath,bbm}
\usepackage{graphicx}
\usepackage{epsfig}
\usepackage{latexsym}
\usepackage{mathtools}
\usepackage{hyperref}
\usepackage[vcentermath]{youngtab}
\hypersetup{
    colorlinks=true,
   linkcolor=black,
    citecolor=black,
    filecolor=black,
    urlcolor=black,}
    
\newcommand{\pathtodiagrams}{figures/}

\newcommand{\mathfig}[2]{{\hspace{-3pt}\begin{array}{c}%
  \raisebox{-2.5pt}{\includegraphics[width=#1\textwidth]{\pathtodiagrams #2}}%
\end{array}\hspace{-3pt}}}

\def \be  {\begin{equation}}
\def \ee  {\end{equation}}
\def \bea {\begin{equation}\begin{aligned}}
\def \eea {\end{aligned}\end{equation}}
\def \ba  {\begin{eqnarray}}
\def \ea  {\end{eqnarray}}
\def \bb  {\begin{thebibliography}}
\def \eb  {\end{thebibliography}}
\def \lab #1 {\label{#1}}

\newcommand\cF{\mathcal{F}}
\newcommand\cG{\mathcal{G}}

\newcommand\cN{\mathcal{N}}
\newcommand\cO{\mathcal{O}}

\newcommand\cS{\mathcal{S}}

\newcommand\cV{\mathcal{V}}
\newcommand\cW{\mathcal{W}}

\newcommand\cZ{\mathcal{Z}}
\newcommand\al{\alpha}

\newcommand\q{\mathfrak{q}}
\newcommand\af{\mathfrak{a}}
\newcommand\tf{\mathfrak{t}}

\newcommand\lb{\lambda}

\newcommand\la{\langle}
\newcommand\ra{\rangle}
\newcommand\del{\partial}
\newcommand\delbar{\bar{\partial}}

\newcommand\tr{\mathrm{Tr}}

\title{Defect Networks and Supersymmetric Loop Operators}
\author{Mathew Bullimore}
\affiliation{Perimeter Institute for Theoretical Physics, \\ Waterloo, Ontario\,N2L 2Y5, Canada}

\abstract{We consider topological defect networks with junctions in $A_{N-1}$ Toda CFT and the connection to supersymmetric loop operators in $\cN=2$ theories of class $\cS$ on a four-sphere. Correlation functions in the presence of topological defect networks are computed by exploiting the monodromy of conformal blocks, generalising the notion of a Verlinde operator. Concentrating on a class of topological defects in $A_2$ Toda theory, we find that the Verlinde operators generate an algebra whose structure is determined by a set of generalised skein relations. These relations encode the representation theory of a quantum group. In the second half of the paper, we explore the dictionary between topological defect networks and supersymmetric loop operators in the $\cN=2^*$ theory by comparing to exact localisation computations. In this context, the the generalised skein relations are related to the operator product expansion of loop operators.}

\begin{document}
\maketitle

\section{Introduction}
\label{Section:Introduction}

We study a correspondence between non-local observables both in four dimensional gauge theories and in two-dimensional conformal field theories. On the four-dimensional side, we consider supersymmetric loop operators such as Wilson loops, 't Hooft loops and dyonic loops in $\cN=2$ supersymmetric gauge theories on a four-sphere or ellipsoid. On the two-dimensional side, we study a class of topological defect networks in $A_{N-1}$ Toda conformal field theory.

The starting point is a rich class of four-dimensional $\cN=2$ superconformal field theories, obtained by compactifying the superconformal $(2,0)$ theory in six dimensions of type $A_{N-1}$ on a Riemann surface $C$ with punctures~\cite{Gaiotto:2009we,Gaiotto:2009hg}.  The marginal deformations are encoded in the complex structure of $C$, while flavour symmetries and mass deformations are encoded in the punctures. An example is the $\cN=2^*$ theory, consisting of an $\mathrm{su}(N)$ vectormultiplet and a hypermultiplet in the adjoint representation. This theory corresponds to a torus with a simple puncture. In particular, the holomorphic gauge coupling is the complex structure parameter of the torus and S-duality transformations are identified with the mapping class group. A simple puncture encodes a $\mathrm{u}(1)$ flavour symmetry acting on the hypermultiplet and the hypermultiplet mass.

The expectation values of supersymmetric observables in such $\cN=2$ theories on a four-sphere are captured by correlation functions in a conformal field theory on the Riemann surface $C$. The foundational example is the four-sphere partition function. This partition function be computed exactly by localisation when there is lagrangian description of the theory~\cite{Pestun:2007rz} (the full correspondence involves the partition function on an ellipsoid, which was constructed and computed recently in~\cite{Hama:2012bg}). For theories of type $A_1$, the partition function is captured by a certain correlation function of Virasoro primary fields in Liouville theory~\cite{Alday:2009aq}. This was subsequently extended to higher rank with correlation functions in $A_{N-1}$ Toda theory~\cite{Wyllard:2009hg}.

The correspondence is enriched by inserting supersymmetric defects on the four-dimensional side, such as loop operators, surface defects and domain walls. In this paper, we consider supersymmetric loop operators, such as Wilson loops, 't Hooft loops and particularly dyonic loop operators. For example, consider the $\cN=2^*$ theory with gauge algebra $\mathfrak{su}(2)$. Ignoring conditions of mutual locality, the loop operators in this case are labelled by a pair of integers $(e,m)$ modulo the identification~\cite{Kapustin:2005py}
\be
(e,m) \sim (-e,-m)\, .
\ee
We can always choose $m\geq 0$ and $e\geq0$ whenever $m=0$. Then $(e,m)$ are interpreted respectively as the electric and magnetic charge of the loop operator. Remarkably, this data is in one-to-one correspondence with homotopy classes of closed curves without intersections on a torus~\cite{Drukker:2009tz}. This observation can be made precise in the context of Liouville theory: the expectation value of a supersymmetric loop operator is given by a correlation function in the presence of topological defects supported on a collection of closed curves without intersection~\cite{Alday:2009fs,Drukker:2009id}. 

Correlation functions in the presence of topological defects can be conveniently expressed in terms of Verlinde operators, which are operators acting on the space of Virasoro conformal blocks. They depend only on the homotopy class of a closed curve $\gamma$ and an irreducible representation of the Virasoro algebra, which in the present case is a degenerate representation with momentum $\al = -b/2$. Roughly speaking, the Verlinde operator measures the monodromy of the conformal blocks obtained by transporting the chiral primary with momentum $\al = -b/2$ around the curve $\gamma$. It has been shown that, acting inside a correlation function, each Verlinde operator introduces a topological defect as constructed in boundary conformal field theory by the unfolding trick~\cite{Drukker:2010jp,Petkova:2009pe}. 

However, for theories of type $A_{N-1}$ with $N>2$, the spectrum of supersymmetric loop operators is much richer. For example, supersymmetric loop operators in the $\cN=2^*$ theory are labelled more generally by an electric and magnetic weight $(\lb_e,\lb_m)$ modulo Weyl transformations,
\be
(\lb_e,\lb_m) \sim (w(\lb_e),w(\lb_m)) \qquad w\in S^N\, .
\ee 
A subset of loop operators that lie in the S-duality orbit of a Wilson loop in the fundamental representation have been matched to Verlinde operators / topological defects supported on closed curves in $A_{N-1}$ Toda theory~\cite{Drukker:2010jp,Gomis:2010kv,Passerini:2010pr}. However, closed curves without intersections cannot account for the complete spectrum of loop operators. 

The exact matches found in the literature have lead to the suggestion that supersymmetric loop operators should correspond in general to topological defect networks with junctions in $A_{N-1}$ Toda theory on $C$~\cite{Drukker:2010jp,Gomis:2010kv}. The expectation that UV line operators in theories of class $\cS$ are labelled by networks has also been explored at a classical level in~\cite{Xie:2013lca} by exploiting the expected relationship to certain coordinate systems on the moduli space of local systems on $C$.\cite{Xie:2013lca}

\begin{figure}[ht]
\centering
\includegraphics[height=2.5cm]{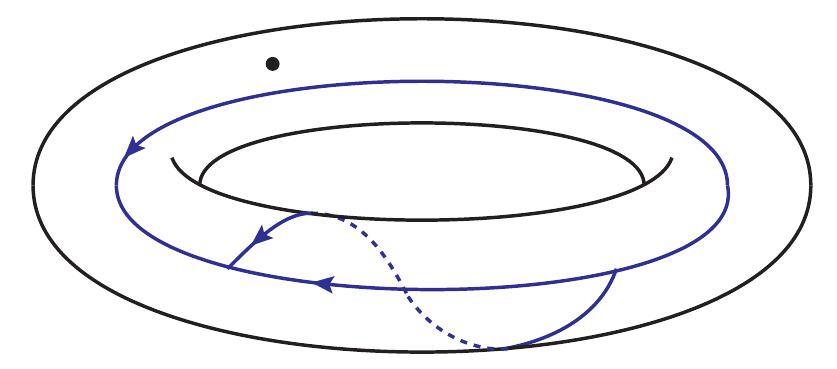}
\caption{\textit{An example of a defect network in $A_2$ Toda theory on a torus with simple puncture. It corresponds to the dyonic loop operator labelled by the weights $(-\omega_2,\omega_1)$ in the $\cN=2^*$ theory.}}
\label{fig:example}
\end{figure}

The aim of this paper is to propose and generate modest evidence for an exact one-to-one correspondence between supersymmetric loop operators in theories of class $\cS$ on a four-sphere and a class of topological defect networks in $A_{N-1}$ Toda theory. The defect lines in this class are oriented and labelled by  $j =1,\ldots,N-1$ such that inverting the orientation is equivalent to interchanging $j \to N-j$. The topological defect lines can end on three-valent junctions that are formed whenever $i+j+k=N$ for incoming lines. The basic ingredients are thus summarised below:
\be
\mathfig{0.4}{diagrams/conjugate} \qquad\qquad \mathfig{0.2}{diagrams/intertwiner}
\ee
For Liouville theory, $N=2$, the defect lines are unoriented and there are no vertices, so we recover closed curves without intersections. In the present paper, we focus on the case of $N=3$. In this case, any line labelled by `2' can be replaced with one labelled by `1' by reversing the orientation. Thus we can drop the labels and consider oriented lines ending on junctions with three incoming or three outgoing lines. An example of such a topological defect in the $\cN=2^*$ theory corresponding to a dyonic loop operator is shown in figure~\ref{fig:example}.

Let us now summarise the contents of each section. In section~\ref{Section:Toda}, we construct topological defect networks by generalising the construction of Verlinde operators. Each defect line labelled by $j$ corresponds to transporting a chiral primary with degenerate momentum $\al = -b\omega_j$ where $\omega_j$ is $j$-th fundamental weight of $A_{N-1}$. The Verlinde operator corresponding to a given network is formed by transporting chiral primaries along each curve in the network and fusing them at the junctions. 

Via this construction, we find that Verlinde operators can often be simplified by removing certain contractible networks. For example, in Liouville theory a disconnected contractible loop can be removed by
\be
\mathfig{0.08}{defects/contractible_loop_A1} = \q+\q^{-1} 
\ee
where $\q=e^{i\pi b^2}$ and $b$ is the dimensionless coupling of Liouville / Toda theory. For Verlinde operators in $A_2$ Toda theory, we find more complicated system of rules for removing contractible networks,
\bea
\mathfig{0.08}{defects/contractible_loop_A2} & = \, \q^2 + 1 + \q^{-2} \\
\mathfig{0.15}{defects/contractible_bubble} & = \, -(\q+\q^{-1}) \, \mathfig{0.13}{defects/line}
\\
& \hspace{-1.87cm} \mathfig{0.47}{defects/contractible_square}\, .
\eea
Let us call Verlinde operators / topological defects where all such contractions have been performed \emph{irreducible}. The conjecture is then that supersymmetric loop operators are in 1-1 correspondence with irreducible topological defect networks in Toda conformal field theory of type $A_{N-1}$.

An important problem is to understand the algebra generated by the Verlinde operators under composition and addition. We find that the composition of two Verlinde operators can always be decomposed by superimposing the corresponding networks and resolving the crossings according to a set of generalised skein relations. The result can then be simplified by the contractions above. For Liouville theory, the skein relations are~\cite{Drukker:2009id}:
\be
\mathfig{0.08}{defects/skein_abelian_2} = \, \q^{1/2} \mathfig{0.08}{defects/skein_abelian_3} + \, \q^{-1/2} \mathfig{0.08}{defects/skein_abelian_4}
\ee
At higher rank, the generalised skein relations involve Verlinde operators with junctions. For $N=3$, we will derive the following generalised skein relations
\bea
& \mathfig{0.09}{defects/skein_2} = \q^{-2/3} \,\mathfig{0.09}{defects/skein_3} +  \q^{1/3}  \mathfig{0.09}{defects/skein_4} \\
& \mathfig{0.09}{defects/skein_2} =  \q^{2/3} \mathfig{0.09}{defects/skein_3} +  \q^{-1/3} \mathfig{0.09}{defects/skein_4} 
\eea
Once the dictionary between supersymmetric loop operators and networks has been constructed, the skein relations allow the operator product expansion of the corresponding loop operators to be computed. 

The removal of contractible loops and the generalised skein relations for Liouville theory and $A_2$ Toda theory are isomorphic respectively to the `spider' relations~\cite{Kuperberg:kx}. These relations encode the representation theory of the quantum groups $U_{\q}(\mathrm{sl}(2))$ and $U_{\q}(\mathrm{sl}(3))$ respectively with $\q=e^{i\pi b^2}$ and are important in the construction of quantum group invariants of knots. Perhaps this connection is not unexpected given the relationship between Liouville / Toda theory and an analytic continuation of Chern-Simons theory~\cite{Verlinde:1989ua,Gaiotto:2011nm}. Therefore, we tentatively conjecture that the algebra of Verlinde operators for irreducible networks in $A_{N-1}$ Toda theory is equivalent to the higher rank spiders constructed in~\cite{Kim:2006fk,Morrison:2007uq}.

\medskip

In section~\ref{Section:N=2*}, we construct the Verlinde operators in a concrete example: $A_2$ Toda theory on a torus with a simple puncture. By extending the exact field theory computation~\cite{Gomis:2011pf} of the expectation value 't Hooft loops in $\cN=2^*$ to the case of dyonic loop operators, we construct the dictionary between supersymmetric loop operators ( in cases without monopole bubbling ) and Verlinde operators for irreducible networks. Furthermore, we compare the generalised skein relations to the expected operator product expansion of loop operators. For loop operators supported on Hopf linked circles, we also find a set of generalised 't Hooft commutation relations and the conditions for mutual locality.

Finally, in section~\ref{Section:Discussion} we discuss what could not be achieved in the present paper and some directions for further research. We include four appendices containing additional material and computational details.



\section{Topological Defects Networks in Toda CFT}
\label{Section:Toda}

\subsection{Review of Toda CFT}

Let us briefly review some points about Liouville / Toda conformal field theory of type $A_{N-1}$. A more thorough introduction and discussion of what is known can be found for example in~\cite{Teschner:2001rv,Fateev:2005gs,Fateev:2007ab,Fateev:2008bm} whose notation we generally follow. Our group theory conventions are summarised in appendix~\ref{appendix:group}. 

Toda conformal field theory is constructed from a two-dimensional scalar field $\varphi$ valued in the Cartan subalgebra of $A_{N-1}$. The case of $N=2$ corresponds to Liouville theory. Introducing a background metric $g$ with scalar curvature $R$, the action functional can be written as
\be
S = \int_X d^2 x \sqrt{g} \, \left(  \frac{1}{8\pi} g^{ab} ( \del_a \varphi ,\del_b \varphi) + \frac{(Q,\varphi)}{4\pi} R   +\mu \sum_{j=1}^{N-1}e^{ b( e_j,\varphi)} \right) \, .
\label{eq:todaaction}
\ee
where $b \in \mathbb{R}$ is a dimensionless parameter and $\mu$ is a dimensionful scale parameter. The scalar $\varphi$ is coupled to the curvature with charge 
\be
Q=q \, \rho \qquad  q \equiv b+b^{-1}\, .
\ee 
The semiclassical limit is defined by rescaling  $\varphi \to \varphi / b$ and taking the limit $b\to0$ with $\lambda \equiv 2\pi \mu b^2=1$ fixed. In this limit, the path integral based on the action~\eqref{eq:todaaction} is dominated by solutions to the classical Toda equations with appropriate boundary conditions. However, interpreting the semiclassical limit of correlation functions in terms of a path integral is rather subtle~\cite{Harlow:2011ny}. The interpretation of topological defect networks in the classical theory is outlined in appendix~(\ref{Appendix:Classical}) and will not be discussed further in the main text.


In practice, correlation functions are constructed by exploiting symmetries of the theory. The chiral algebra of Toda conformal field theory is generated by holomorphic currents $W_j(z)$ of spin $j=2,\ldots,N$ and is known as the $W_N$--algebra. In particular, the holomorphic stress tensor
\be
W_2(z) = (Q, \del^2 \varphi ) - \frac{1}{2} (\del \varphi, \del \varphi)
\ee
generates a Virasoro subalgebra with central charge $c=N-1+12q^2$. The irreducible representations of the $W_N$-algebra are labelled by a complex vector $\al$ in the Cartan subalgebra of $A_{N-1}$ and the corresponding primary fields are exponentials
\be
V_{\al} = e^{(\al,\varphi)}\, .
\ee
The representations are characterised by charges $w_j(\al)$ under the holomorphic currents. In particular, the conformal dimension is given by
\be
\Delta(\al)=\frac{(2Q-\al,\al)}{2}\, .
\ee 
These charges $w_j(\al)$ are invariant under Weyl transformations
\be
s : \al \mapsto  Q + s ( \al - Q ) \qquad s \in S_N
\ee
and an important fact is that correlation functions of the corresponding primary fields are proportional. An example is that the vertex operator with momentum $2Q-\al$ is proportional to that with momentum $\al^*$, where conjugation is defined by $(\al^*,e_j)=(\al,e_{N-j})$. 

The complete symmetry algebra of the theory on a cylinder is given by holomorphic and anti-holomorphic copies of the $W_N$--algebra, and the Hilbert space is defined by the diagonal pairing of representations
\be
\int d \al \;  \cV_{\al} \otimes \cV_{\al}
\ee
where $\al = Q +ia$ with $a\in \mathbb{R}^{N-1}$ are non-degenerate representations corresponding to delta-function normalisable states. In addition, there exists semi-degenerate and completely degenerate representations of the $W_N$--algebra that play a role in what follows. The spectrum of such representations and their characters are discussed, for example, in~\cite{Drukker:2010jp}.  

Of particular importance here are completely degenerate representations with primary operators with momentum $\al = -b\lambda$ where $\lambda$ is a dominant integral weight. They are thus labelled by irreducible representations of $A_{N-1}$. The operator product expansions of the corresponding primary fields are simple
\bea
V_{-b \lambda} \cdot V_{\al} & = \sum_{j} V_{\al - b \, w_j} \\
V_{-b\lambda_1} \cdot V_{-b\lambda_1} &= \sum_{\lambda_3} \cN_{\lambda_1 \lambda_2}^{\lambda_3} V_{-b\lambda_3}
\label{degenerateope}
\eea
where $w_j$ are weights of the representation with highest weight $\lambda$ and $\cN_{\lambda_1 \lambda_2}^{\lambda_3}$ are the Littlewood-Richardson coefficients. Thus the operator product expansion mirrors the structure of the representation theory of $A_{N-1}$. 

\subsection{Topological Defect Networks}

Topological defects are important observables in any two-dimensional conformal field theory. Although, we concentrate here in a particular class of topological defects in Liouville / Toda conformal field theory, much of the construction that follows could be framed more generally, for example, in the context of rational conformal field theories.

\begin{figure}[h]
\centering
\includegraphics[height=2cm]{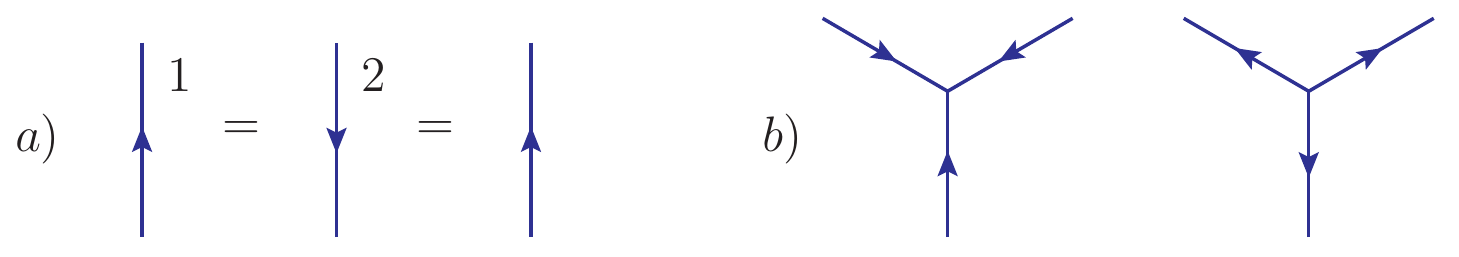}
\caption{\textit{The ingredients for topological defect networks in Toda theory of type $A_2$ are a) oriented lines and b) incoming and outgoing vertices.}}
\label{defectnetwork}
\end{figure}

Here, we are going to study a class of topological defects in Toda conformal field theory of type $A_{N-1}$ supported on closed networks without intersections embedded in a Riemann surface. The networks are constructed from the components:
\begin{itemize}
\item Oriented curves labelled by an irreducible representation of $A_{N-1}$ with fundamental highest weight $\omega_j$ for $j=1,\ldots,N-1$. Inverting the orientation of a line is equivalent to conjugation of the representation.
\begin{equation*}
\mathfig{0.41}{diagrams/conjugate}
\end{equation*}
\item Trivalent vertices whenever there is an invariant tensor on the tensor product of representations $\Lambda_i \times \Lambda_j \times \Lambda_k$, that is whenever $i+j+k=N$.
\begin{equation*}
\mathfig{0.22}{diagrams/intertwiner}
\end{equation*}
\end{itemize} 
For $A_1$, the curves are labelled by the fundamental representation $\Lambda_1$ and so are unoriented. There are no vertices, and the topological defects are constructed from unions of closed non-self-intersecting curves. For our main example, $A_2$, we have oriented curves labelled by the fundamental $\Lambda_1$ and anti-fundamental $\Lambda_2$ representations. Since reversing the orientation interchanges the labels $\Lambda_1\leftrightarrow\Lambda_2$, it is convenient to transform all lines to the fundamental and then omit the label. There is then only one vertex and its conjugate - see figure~\ref{defectnetwork}.

From each network of (together with a choice of `framing' that we discuss later), there is a recipe to construct an operator acting on the space of $W_N$-algebra conformal blocks in a way that generalises the notion of a Verlinde loop operator~\cite{Verlinde:1988sn}. The operator is defined by exploiting the monodromy properties of conformal blocks under fusion and braiding of the degenerate primary fields with momenta $\al = -b\omega_j$. The operator constructed in this fashion is invariant under smooth deformations of the network. 

The starting point for this construction is the expansion of a given correlation function into Liouville / $W_N$--algebra conformal blocks. A correlator of $n$ primary fields $V_{\mu_1}, \ldots, V_{\mu_n}$ can be computed by picking a pants decomposition $\sigma$ of the Riemann surface $X$. Each pants decomposition can be represented by a trivalent graph $\Gamma_{\sigma}$ that encodes a sewing of the Riemann surface from $(2g-2+n)$ pairs-of-pants and $(3g-3+n)$ tubes. For each pants decomposition $\sigma$ there is a set of Virasoro / $W_N$--algebra conformal blocks
\be
\cF^{(\sigma)}_{\al,\mu}(\tau)
\ee
depending on
\begin{enumerate}
\item Representations on external edges: $\mu=(\mu_1,\ldots,\mu_n)$.
\item Representations on internal legs: $\al = (\al_1,\ldots,\al_{3g-3+n})$.
\item Sewing parameters on each edge: $\tau = (\tau_1,\ldots,\tau_{3g-2+n})$.
\end{enumerate}
The conformal blocks also depend on a choice of fusion channel at each vertex, whose dependence we are suppressing. The correlator can then be expanded
\be
\la V_{\mu_1} \ldots V_{\mu_n} \ra  = \int_{\al} d \nu(\al) \, \bar{\cF}^{(\sigma)}_{\al,\mu}(\tau) \, \cF^{(\sigma)}_{\al,\mu} (\tau)\,
\ee
where the integral is over all internal representations and fusion channels allowed by the operator product expansion of primary fields. The measure $\nu(\al)$ contains an appropriate three-point function for each vertex of the graph $\Gamma_\sigma$. Associativity of the operator product expansion ensures that the whole expression is independent of the choice of $\sigma$. 

Now, a Verlinde operator supported on an embedded network can be computed by a sequence of fusion and braiding moves on the $W_N$-algebra conformal blocks:
\begin{enumerate}
\item Insert the identity operator $\mathbbm{1}$ at some point on the trivalent graph $\Gamma_\sigma$.
\item Resolve the identity operator $\mathbbm{1}$ as the fusion of two degenerate chiral primaries $V_{\mu}$ and $V_{\mu^*}$ where the momentum $\mu = -b\omega_i$ corresponds to the label $i$ on a curve $\gamma$ of the network.
\item Transport the primary field $V_{\mu}$ with momentum $\mu=-b\omega_i$ around the trivalent graph $\Gamma_{\sigma}$ according to the orientation of the curve $\gamma$.
\item When the curve $\gamma$ ends on a vertex with three incoming curves $(\gamma,\gamma',\gamma'')$ labelled by integers $(i,j,k)$ such that $i+j+k=N$, resolve the primary $V_{\mu}$ as the fusion of two chiral primaries $V_{\mu'}$ and $V_{\mu''}$ with momenta $\mu'=-b\omega_{N-j}$ and $\mu''=-b\omega_{N-k}$\,.
\item Repeat steps 3. and 4. according to the structure of the network. The endpoint is again the chiral primaries $V_{\mu}$ and $V_{\mu^*}$ in the same configuration as in step 2.
\item Fuse the vertex operators $V_\mu$ and $V_{\mu^*}$ back to the identity operator $\mathbbm{1}$ to close the network.
\end{enumerate}

\noindent This construction leads to a sequence of elementary fusion and braiding operations, whose endpoint is an operator $\cO$ acting the space of Virasoro / $W_N$--algebra conformal blocks subordinate to the original pants decomposition $\sigma$,
\bea
&\cF^{(\sigma)}_{\al,\mu} \to \big[ \cO \cdot \cF^{(\sigma)} \big]_{\al,\mu}  = \sum_{\al'} \cO_{\al,{\al'}} \, \cF^{(\sigma)}_{\alpha',\mu} \\
&\overline\cF^{(\sigma)}_{\al,\mu} \to  \overline\cF^{(\sigma)}_{\al,\mu}\, \, .
\eea
where $\cO_{\al\al'}$ are the matrix elements of the operator. Note that due to the operator product expansion of degenerate primary operators~\eqref{degenerateope}, the Verlinde operator always takes the form of a discrete sum of terms. There are natural operations of summation and composition of Verlinde operators defined in the obvious way. An important goal is to understand the algebra generated by the above operations, that is, how to decompose a composition $\cO_1 \circ \cO_2$ as a sum of operators $\sum_j c_j \cO_j$. Below, we show that this is found by superimposing the corresponding networks and resolving each crossing according to a set of generalised skein relations.

\begin{figure}[h]
\centering
\includegraphics[height=4.5cm]{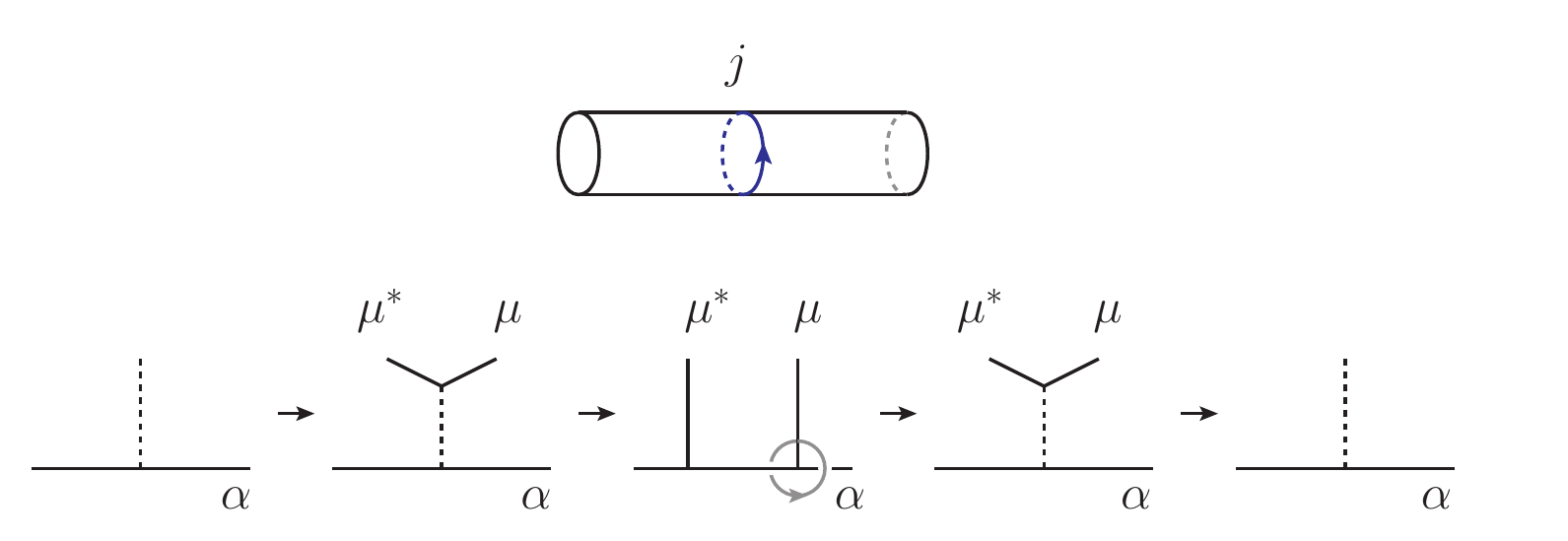}
\caption{\textit{A topological defect a) labelled by the representation with momentum $\mu=-b\omega_j$ wrapping the tube of a pants decomposition with momentum $\al$, and the sequence b) of fusion and braiding operations needed to compute the Verlinde operator.}}
\label{fig:cardy}
\end{figure}

The Verlinde operator itself depends on the choice of pants decomposition $\sigma$ by definition.  However, acting with a Verlinde operator inside a correlation function produces a result that is independent of the choice of pants decomposition $\sigma$,
\be
\int d\nu(\al) \, \overline\cF^{(\sigma)}_{\al,\mu} (\tau) \big[\, \cO \cdot \cF^{(\sigma)} \big]_{\al,\mu} (\tau) \, \equiv \,  \la \, \cO \,  \ra
\ee
and defines the expectation value of a topological defect operator in Toda conformal field theory. To see this, let us consider the Verlinde operator corresponding to a closed curve labelled by momentum $\mu=-b\omega_j$ wrapping a tube of the pants decomposition $\sigma$ with momentum $\al$. This can be computed by the sequence of operations illustrated in figure. Combining this sequence of operations leads to a multiplicative operator
\be
\big[ \cO \cdot \cF^{(\sigma)} \big]_{\al,\ldots} = \frac{S_{\mu,\al}}{S_{1,\al}} \, \cF^{(\sigma)}_{\al,\ldots}
\ee
where $S_{\mu,\al}$ denotes the modular S-matrix~\cite{Drukker:2010jp}. Inside a correlation function, this agrees with expectation value of a topological defect labelled by the representation of the $W_N$-algebra with momentum $\mu=-b\omega_j$, as defined in boundary conformal field theory by the unfolding trick. When there are junctions, a classification in terms of boundary conformal field theory cannot be applied. Nevertheless, we expect that the Verlinde operators considered here coincide with topological defect networks defined axiomatically, for example in~\cite{Frohlich:2006ch}. 

Finally, the Verlinde operator $\cO$ depends on a choice of framing of the network. This modifies the operator by fractional powers of the quantum parameter $\q = e^{i\pi b^2}$. This ambiguity disappears in the semi-classical limit $b\to0$. In practice, we will make choice for a subset of networks that are sufficient to generate the algebra, and then impose a choice of quantum Skein relations. This choice is designed to be compatible with the 't Hooft commutation relations of loop operators in four-dimensional gauge theories.

\subsection{Building Blocks}

In this section, we compute some of the basic ingredients needed to compute the aforementioned class of defects in the case of $A_2$ Toda theory. The ingredients that we can compute are all obtained by exploiting monodromy properties of conformal blocks contributing to the four-point correlation function shown in figure. This correlator was computed in~\cite{Fateev:2005gs,Fateev:2007ab} and the monodromy properties of its conformal blocks are reviewed extensively in appendices~(\ref{appendix:hypergeometric}) and~(\ref{appendix:fusion}). 

\begin{figure}[h]
\centering
\includegraphics[height=2.5cm]{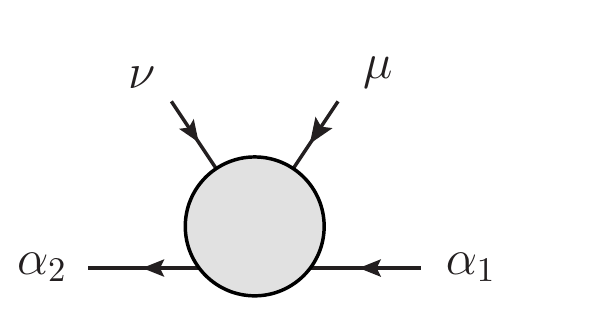}
\caption{\textit{The four-point correlation function of with two non-degenerate momentum $\al_1$ and $\bar\al_2 = 2Q-\al_2$, one semi-degenerate momentum of the form $\nu\equiv-\kappa h_N$ where $\kappa\in \mathbb{R}$, and one completely degenerate momentum $\mu\equiv-bh_1$. In what follows we omit the arrows on conformal block diagrams, with the understanding that they are pointing downwards and to the right, as shown here.}}
\label{4point_generic}
\end{figure}

Introducing the convenient parametrisation
\be
\la V_{2Q-\al_2}(\infty)V_{\nu}(1)V_{\mu}(z) V_{\al_1}(0) \ra = |z|^{-2 b (\mu-Q,h_1)}|1-z|^{2b(\nu,h_1)} G(z,\bar{z})\, ,
\label{4point}
\ee 
the function $G(z,\bar{z})$ is constructed from diagonal combinations of solutions to an n-th order generalised hypergeometric differential equation
and its conjugate with $z\leftrightarrow\bar{z}$. The hypergeometric equation has regular singularities at points $z=0,1,\infty$ corresponding to the three pants decompositions of a sphere with four-punctures: the s-channel, t-channel and u-channel respectively. Solutions to the hypergeometric equation with diagonal monodromy around the singular points provide a basis of conformal blocks for this four point function.

The conformal block for a particular representation in the intermediate channel can normally be identified by its monodromy, which is dictated by the operator product expansion. In the s- and u-channels, we have
\be
\begin{aligned}
V_{\mu}(z) \cdot V_{\al_1}(0) & = \sum_{j=1}^N C_j^{(s)} \left( \, |z|^{2 s_j} V_{\al_1-b h_j}  + \ldots \right) \\
V_{\mu}(z) \cdot V_{\bar\al_2}(\infty) & = \sum_{j=1}^N C_j^{(u)} \left( |z|^{-2 u_j} V_{\bar\al_2+b h_j}  + \ldots \right)
\end{aligned}
\ee
where
\be
\begin{aligned}
& s_j = i b a_{1,j} + \frac{1}{2}bq(N-1) \\
& u_j = -i a_{2,j}-\frac{1}{2}bq(N-1)-b^2\left(1-1/N \right)\, .
\end{aligned}
\ee
Correspondingly, the monodromy matrices at the regular singular points $z=0$ and $z=\infty$ are non-degenerate with eigenvalues $e^{2 \pi i s_j}$ and $e^{2 \pi i u_j }$ respectively. The conformal blocks in the s-channel and u-channel can therefore be therefore uniquely identified by their monodromy. On the other hand, the operator product expansion in the t-channel predicts the appearance of just two representations in the intermediate channel
\be
\begin{aligned}
V_{\mu}(z) \cdot V_{\nu}(1) & = \, C_1^{(t)} \left( \, |1-z|^{2 t_1} \, V_{\nu-bh_N} + \cdots \right) \\
& + \,  C_2^{(t)} \left( \, |1-z|^{2 t_2} \, V_{\nu-b h_1} + \cdots \right)
\end{aligned}
\ee
where
\be
\begin{aligned}
t_1 & = bq(N-1) - b\kappa(1-1/N) \\
t_2 & = b \kappa/N \, ,
\end{aligned}
\ee
The monodromy matrix at the regular singular point $z=1$ is degenerate, with the eigenvalue $e^{2\pi i t_1}$ appearing once and the eigenvalue $e^{2\pi i t_2}$ appearing $(N-1)$ times. The reason for the degeneracy in the monodromy matrix is that some descendants of $V_{\nu-bh_1}$ can appear independently in the conformal block expansion. It is an important feature of Toda conformal field theory that the correlation functions of primary fields are not sufficient to construct the correlators of all $W_N$--algebra descendants.  Nevertheless, in the examples studied below, we find that the correct conformal block can always be isolated from the degenerate subspace by consistency with the operator product expansion in another channel. 

\subsubsection{Fusion 1}

Let us consider the process of creating and annihilating topological defect lines labelled by the fundamental and anti-fundamental representations. This corresponds to resolving the trivial representation by $\mathbbm{1} \subset V_\mu \times V_{\mu^*}$ and then fusing them again. For this subsection, we can keep $n$ generic. 

The required conformal blocks are obtained by restricting the parameter of the semi-degenerate puncture in the above four-point correlator to $\kappa=-b$. The operator product expansion
\be
\begin{aligned}
V_{\mu}(z)\cdot V_{\mu^*}(1) & \, = \, C_{\mu,\mu^*}^{0} \left( |1-z|^{-4\Delta(\mu)} \mathbbm{1} + \cdots \right) + \cdots \\
\end{aligned}
\ee
where
\be
2 \Delta(\mu)=-b^2(N-1/N)-(N-1)
\ee
dictates that the t-channel conformal block with the trivial representation in the intermediate channel has monodromy $\q^{2(N-1/N)}$ around $z=1$. From appendix, the unique conformal block with this monodromy can be expressed as a linear combination of s-channel conformal blocks
\vspace{-1mm}
\be
\mathfig{0.20}{blocks/stchannel-trivial-1} \hspace{-8mm}=  \frac{\Gamma(Nbq)}{\Gamma(bq)} \sum_{k=1}^N  \Bigg[ \; \prod\limits_{j\neq k}^N \, \frac{\Gamma\left(iba_{jk} \right)}{\Gamma(iba_{jk}+bq)}  \, \mathfig{0.24}{blocks/stchannel-trivial-2} \hspace{-4mm}\Bigg]
\ee
\vspace{-5mm}

\noindent where $\Gamma(x)$ is the Euler gamma function. The corresponding annihilation process is obtained by extracting the unique term in the inverse transformation with the monodromy eigenvalue $e^{2\pi i b^2(N-1/N)}$. The result is the projection
\vspace{-2mm}
\be 
 \mathfig{0.24}{blocks/stchannel-trivial-2} \hspace{-8mm} \longrightarrow \; \frac{\Gamma(1-Nbq)}{\Gamma(1-bq)} \Bigg[ \; \prod_{j\neq k}  \frac{\Gamma\left( 1-iba_{jk}\right)}{\Gamma(1-ia_{jk}-bq)} \;  \Bigg]\; \mathfig{0.20}{blocks/stchannel-trivial-1} \, .
\ee
\vspace{-5mm}

\subsubsection{Fusion 2}

In Liouville theory, the fundamental representation of $A_1$ is self-conjugate and so there are no vertices involving only the fundamental representation $V_{\mu}$. Let us consider the next simplest case, $A_2$ Toda theory. There are two vertices connecting defect lines in the fundamental and anti-fundamental representations, corresponding to the fusions  $V_{\mu^*} \subset V_{\mu} \times V_{\mu}$ and $V_{\mu} \subset V_{\mu^*} \times V_{\mu^*}$. Therefore, we require the monodromy properties of conformal blocks for the four-point correlator of two vertex operators with degenerate momentum $\mu=-bh_1$ and two further vertex operators with non-degenerate momenta. In this case, the required correlator can be obtained from the general four-point correlator of figure~(\ref{4point_generic}) by performing a reflection of the semi-degenerate momentum $\nu=-\kappa h_3$ and then specialising to a particular value of the real parameter $\kappa$. 

\begin{figure}[h]
\centering
\includegraphics[height=2.5cm]{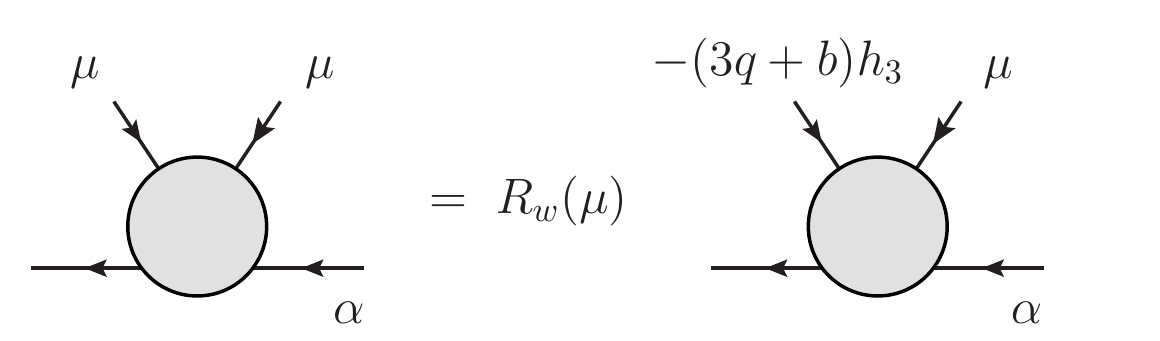}
\caption{\textit{In $A_2$ Toda theory, the four-point correlator with two primary fields of equal degenerate momentum $\mu=-b\omega_1$ is related to that standard four-point function with the specialisation $\kappa = 3q+b$ by applying the Weyl transformation $w:h_j \to h_{j+1}$. The conformal blocks of the above four-point correlators are therefore equal.}}
\label{fig:4point_trick}
\end{figure}

The Weyl group of $A_{N}$ consists of permutations of the weights $\{h_1,\ldots,h_N\}$ of the fundamental representation. An element $w$ acts on the momentum of a representation by $\al\to Q+w(\al-Q)$, leaving invariant the conformal dimension $\Delta(\al)$ and the higher spin charges. An important fact is that correlation functions of the corresponding vertex operators are proportional
\be
V_{\al} = R_w(\al) \, V_{Q+w(\al-Q)}
\ee
where the coefficient $R_{w}(\al)$ is known as the reflection amplitude (see~\cite{Fateev:2007ab} for the explicit form of the reflection amplitude). Since this is an operator relation, it holds for the three-point functions of the theory. This means that given a pants decomposition of a higher point correlator, the reflection amplitude is always absorbed into a three-point function. Thus, the $W_N$-algebra conformal blocks are invariant under Weyl transformations of the momenta.

An important example is the transformation $w:\{h_1 \leftrightarrow h_N\}$, which sends the momentum $\bar\al = 2Q-\al$ to the conjugate momentum $\al^*$ defined by the inner product with the simple roots $(\al^*,e_j) = (\al, e_{n-j})$. For our present problem in the case of $A_2$, we note that the cyclic permutation $w:\{ h_{j} \to h_{j-1}\}$ applied to a completely degenerate momentum $\mu$ transforms it to
\be
w: -bh_1 \to  - (3q+b) h_3\, .
\ee
Thus, the conformal blocks we need can be obtained by specialising to
\be
\kappa=3q+b
\ee
in the standard four-point function, as illustrated in figure~\ref{fig:4point_trick}

To compute the vertex amplitude from the four-point function, we must isolate the t-channel conformal block corresponding to the fusion $V_{\mu^*} \subset V_{\mu} \times V_{\mu}$ in the intermediate channel. From the relevant operator product expansion
\be
\begin{aligned}
V_{\mu}(z) \cdot V_{\mu}(1) & \, = \, C_{\mu,\mu}^{\mu^*} \left( |1-z|^{2(1+4b^2/3)} V_{\mu^*}  + \cdots \right) + \cdots \, ,
\end{aligned}
\ee
the relevant solution has monodromy $\q^{8/3}$ around $z=1$. However, specialising to $\kappa=3q+b$, we find that the monodromy operator around $z=1$ has one eigenfunction with monodromy $\q^{-4/3}$ and two eigenfunctions with degenerate monodromy $\q^{8/3}$.

To isolate the correct conformal block from the degenerate subspace, we check consistency with the expected representations appearing in the s-channel. For the t-channel conformal block corresponding to $V_{\mu^*}$ in the intermediate state, the external momenta are restricted to
\be
\al_1 = \al \qquad \al_2 = \al+bh_i \qquad i=1,2,3\, .
\ee
Now, transforming to the s-channel, the operator product expansion dictates that the intermediate channel can have momentum $\al-bh_j$ for $j\neq i$. This uniquely determines the correct linear combination of solutions for a given $i$, up to an overall scale that is fixed by demanding the leading term as $z\to 1$ has unit coefficient. They are
\bea
\mathfig{0.23}{blocks/tchannel_splitting_i} \hspace{-8mm} \; &= \; \cN^{(i)}_{j} \; \mathfig{0.26}{blocks/schannel_splitting_ij} \hspace{-8mm} \; - \; \cN^{(i)}_{k} \; \mathfig{0.26}{blocks/schannel_splitting_ik} \hspace{-8mm} \\
\mathfig{0.23}{blocks/tchannel_conj_splitting_i} \hspace{-8mm} \; &=\; \bar\cN^{(i)}_{j} \mathfig{0.26}{blocks/schannel_conj_splitting_ij} \hspace{-8mm} \; - \;\bar\cN^{(i)}_{k} \; \mathfig{0.26}{blocks/schannel_conj_splitting_ik} \hspace{-8mm}
\eea
where the numbers $\{i,j,k\}$ are a cyclic permutation of $\{1,2,3\}$. In the second line we have included the conjugate process for convenience, which is derived from the invariance under simultaneous conjugation of all momenta and the property $\cN^{(i)}_j = \overline \cN^{(i)}_{k}$ of the function
\bea
\cN^{(i)}_{j} = \Gamma(-2b^2) \prod_{k=1}^{3} \frac{\Gamma(iba_{jk}-b^2(\delta_{ik}-1) +1)}{\Gamma(iba_{jk}+1)} 
\, .
\eea
The inverse process is obtained by inverting the transformation and projecting back onto the t-channel block with monodromy $\q^{8/3}$. From the inverse transformations in appendix~(\ref{appendix:fusion}) we find
\bea
\cN^{(i)}_{j} \; \mathfig{0.26}{blocks/schannel_splitting_ij} \hspace{-8mm} \longrightarrow (-1)^\sigma \frac{\sin\pi b^2}{\sin 2 \pi b^2 }\frac{\sin\pi(iba_{kj}+qb)}{\sin\pi(iba_{kj})} \mathfig{0.23}{blocks/tchannel_splitting_i} \hspace{-8mm} \\
\bar\cN^{(i)}_{j} \; \mathfig{0.26}{blocks/schannel_conj_splitting_ij} \hspace{-8mm} \longrightarrow (-1)^\sigma \frac{\sin\pi b^2}{\sin 2 \pi b^2 }\frac{\sin\pi(iba_{jk}+qb)}{\sin\pi(iba_{jk})} \mathfig{0.23}{blocks/tchannel_conj_splitting_i} \hspace{-8mm}
\eea
where $\sigma$ is the signature of $\{i,j,k\}$. In the second line, we have again included the conjugate operation.

\subsubsection{Braiding and Framing}

In each of the fusion operations above, there is the freedom to perform a number of t-channel braiding operations. Each such braiding introduces an additional phase into the Verlinde operator, determined by the conformal dimensions of the operators involved. For example, interchanging the conjugate primary fields $V_{\mu}$ and $V_{\mu^*}$ in the identity fusion channel introduces the phase
\be
e^{i\pi (\Delta(\mu)+\Delta(\mu^*) )} = (-1)^{N-1} e^{i\pi b^2(1/N-N)}\, .
\ee
In pictures,
\vspace{-2mm}
\be
\mathfig{0.20}{blocks/tchannel-braiding-1} \hspace{-5mm} = \, (-1)^{N-1}\q^{1/N-N} \;\mathfig{0.20}{blocks/tchannel-braiding-2} \, .
\ee
\vspace{-2mm}

\noindent In $A_2$ Toda theory, we can also interchange identical primary fields $V_{\mu}$ and $V_{\mu}$ where $\mu = -b\omega_1$ in the fusion channel $V_{\mu^*}$. In summary, the braiding phases for $A_2$ Toda are given by. 
\bea
\mathfig{0.20}{blocks/tchannel-braiding-1} \hspace{-5mm} & = \, \q^{-8/3} \;\mathfig{0.20}{blocks/tchannel-braiding-2} \\
\mathfig{0.20}{blocks/tchannel-braiding-3} \hspace{-5mm} & = \, -\q^{-4/3} \;\mathfig{0.20}{blocks/tchannel-braiding-4} 
\eea
where in the second line the same formula holds with $\mu \leftrightarrow \mu^*$. 

Performing different numbers of such braidings at intermediate stages is a choice of `framing' for the Verlinde operator. This ambiguity means that we should be drawing the curves in the network as ribbons, in order to keep track of the framing. In practice, we are going to fix the framing in some simple examples and then extend this convention systematically by using the Skein relations. Our choice is motivated by the connection to loop operators in four-dimensional gauge theories.


\subsection{Contractible Loops}
\label{sub:Contractible Loops}

\begin{figure}[h]
\centering
\includegraphics[height=4.2cm]{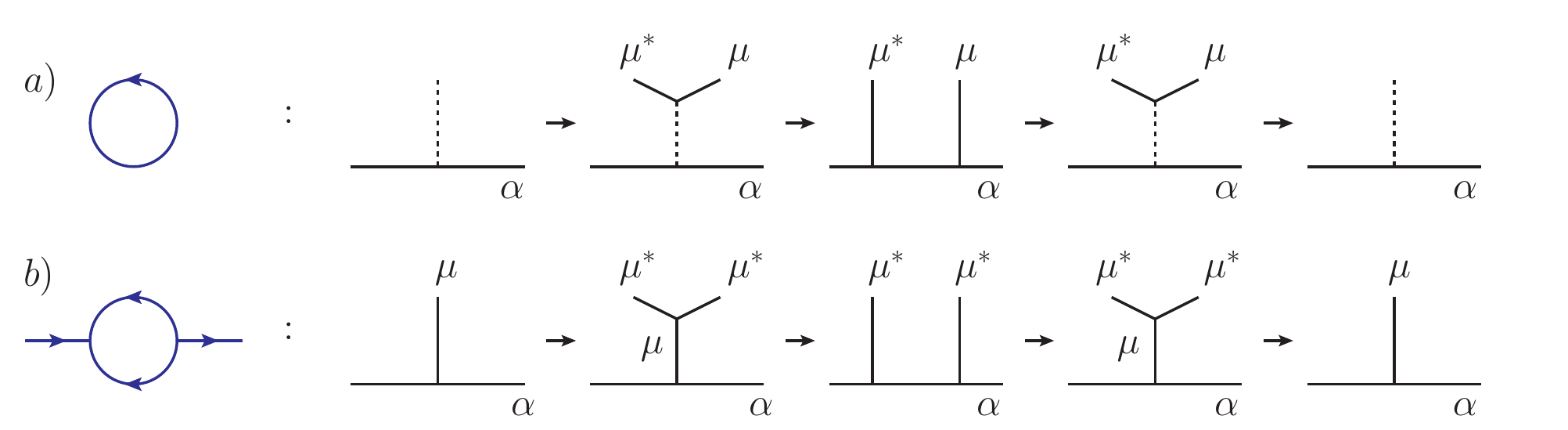}
\caption{\textit{Contractible loop a) and bubble b) in $A_2$ Toda theory and the corresponding sequence of fusion operations.}}
\label{fig:contractible}
\end{figure}

An important consistency condition on the fusion matrices is that topological defects involving certain contractible loops can be simplified by removing the loop. The simplest example for $A_{N-1}$ Toda theory is the creation and immediately annihilation a pair of vertex operators with conjugate momentum $\mu$ and $\mu^*$, which introduces a contractible loop. Combining the relevant fusion matrices according to figure~\ref{fig:contractible} part a) , we find 
\be
\frac{\sin\pi bq}{\sin\pi n bq} \sum_{k=1}^n \prod_{j\neq k} \frac{\sin \pi b( ia_{jk}+q)}{\sin \pi b ( i a_{jk} )} = 1
\ee
Here, we find it convenient to normalise the Verlinde operators to eliminate the prefactor in the above equation. In this normalisation, we associate to each contractible loop the factor
\be
\mathfig{0.1}{defects/contractible_loop} = \frac{\sin n \pi b q}{\sin \pi b q} = (-1)^{n-1}[n]_{\q}\, , 
\label{eq:closedcontloop}
\ee
where the notation $[n]_{\q}=\q^{n-1}+\q^{n-3}+\ldots +\q^{1-n}$ is the quantum dimension of the fundamental representation of $A_{N-1}$ with parameter $\q=e^{i\pi b^2}$. 

In Liouville theory, there are no further independent contractible loops because there are no vertices. However, in $A_2$ Toda theory, starting with momentum $\mu$ and creating and immediately annihilating a pair of vertex operators with identical momentum $\mu^*$, introduces a contractible bubble. Combining the fusion matrices according to the steps in figure~\ref{fig:contractible} part b), we find
\be
\frac{\sin\pi q b}{\sin 2 \pi q b } \left[ \frac{\sin\pi(iba_{kj}+qb)}{\sin\pi(iba_{kj})} + \frac{\sin\pi(iba_{jk}+qb)}{\sin\pi(iba_{jk})}\right] =1\, .
\ee
In the same manner as above, we choose our normalisation to eliminate the prefactor in the above equation and so associate to each contractible bubble a factor
\be
\mathfig{0.2}{defects/contractible_bubble} = \,\frac{\sin 2 \pi q b}{\sin \pi q b} \, \mathfig{0.15}{defects/line} = \, -[2]_{\q} \, \mathfig{0.15}{defects/line}\, .
\label{contractible2}
\ee
Finally, patiently combining the fusion matrices we find that any contractible square graph can be removed by
\be
\mathfig{0.5}{defects/contractible_square}\, .
\label{contractible3}
\ee
It is expected that the above relations generate all possible simplifications that can be made by removing contractible subgraphs of $A_{2}$ networks: contractible $2n$-gons cannot be eliminated for $n>2$. A topological defect network where all contractible loops, bubbles and squares will be called irreducible.


\subsection{Skein Relations}
\label{sub:Skein Relations}

Now, let us consider the composition $\cO_1\circ\cO_2$ of Verlinde operators corresponding to two networks and ask how it can be decomposed into a sum $\sum_j c_j \cO_j$. The result is that one should superimpose the two networks and resolve each crossing according to a set of generalised Skein relations. To derive the Skein relations, let us compute some simple Verlinde operators for networks on a cylinder. We imagine that this configuration is part of a larger network and the cylinder is identified with a tube of the pants decomposition.

\begin{figure}[h]
\centering
\includegraphics[height=3.5cm]{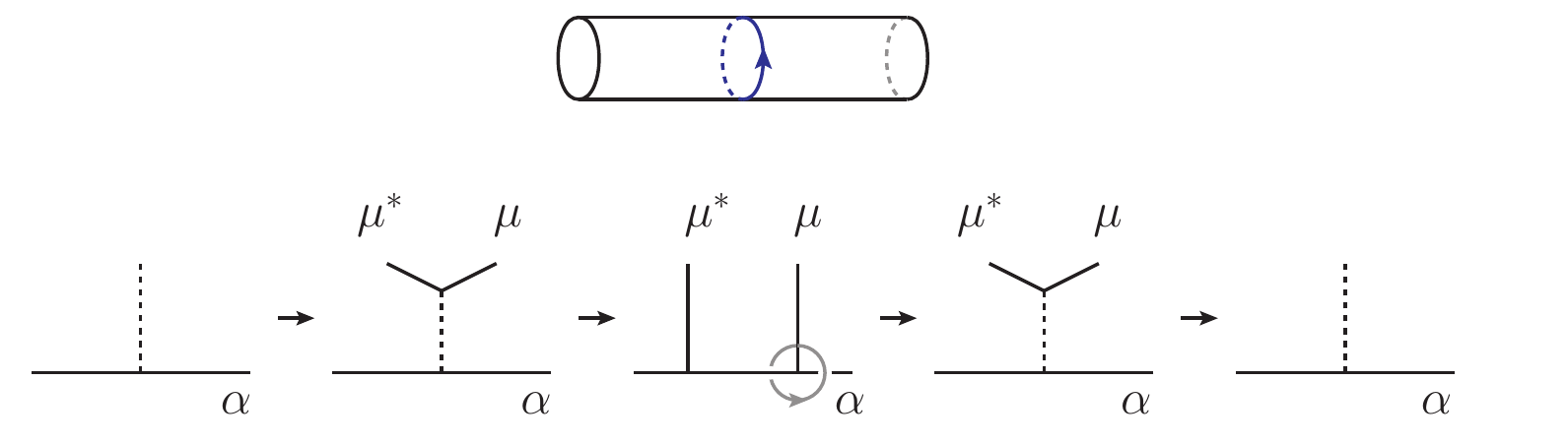}
\caption{\textit{Sequence of fusion and braiding operations for topological defect labelled by momentum $\mu=-b\omega_1$ wrapping the tube of a pants decomposition.}}
\label{fig:wrapping_tube}
\end{figure}

The simplest defect corresponds to a defect line wrapping once around the cylinder. This is computed by the sequence of operations shown in figure~\ref{fig:wrapping_tube}. Multiplying the sequence by the quantum dimension $[n]_{\q}$ we find
\be
e^{i \pi b q (N-1)} \sum_{j=1}^N \prod_{k\neq j} \frac{\sin \pi (iba_{jk}+bq)}{\sin\pi(iba_{jk})} e^{-2\pi b a_j} = \sum_{j=1}^N e^{-2\pi b a_j} \, .
\ee
Note that had we chosen to braid the primary field with the conjugate momentum $\mu^*$ around the tube in the opposite direction, we would find the same result up to an additional phase $\q^{2(1/N-N)}$. Thus, we have made a choice of framing here. Let us introduce the notation $\af_j = e^{-2\pi b a_j}$. Then, braiding the primary with momentum $\mu$ in the opposite direction corresponds to reversing the orientation of the defect line and sends $\af_j \to \af^{-1}_j$. We can summarise the above computation by
\vspace{2mm}
\be
\begin{alignedat}{2}
 \mathfig{0.25}{defects/wilson_loop_p}  &=  \; \sum_{j=1}^N\af_j  \\
 \mathfig{0.25}{defects/wilson_loop_n} &=  \; \sum_{j=1}^N \af^{-1}_j\, .
\end{alignedat}
\ee
\vspace{2mm}

Now consider transporting a chiral vertex operator with degenerate momentum $\mu=-bh_1$ long the tube. From the operator product expansion of degenerate primaries~\eqref{degenerateope}, the Verlinde operator for any network containing this component is a sum operators shifting the momentum associated to the tube,
\be
\begin{alignedat}{2}
\mathfig{0.27}{defects/thooft_loop_p}  &=  \; \sum_{j=1}^N c_j \Delta_j  \\
\end{alignedat}
\ee
where
\bea
\Delta_j &: \al \to \al - b h_j \\
&: \af_i \to \q^{-2(\delta_{ij}-1/N)} \af_i \, .
\eea
The coefficients $c_j$ depend on the momentum $\al$ in the tube and encode the structure on the network outside the tube. In particular, $c_j$ can be a difference operator shifting other internal momenta. Note that reversing the orientation of the defect line would shift the momentum in the opposite direction.

\begin{figure}[h]
\centering
\includegraphics[height=3.5cm]{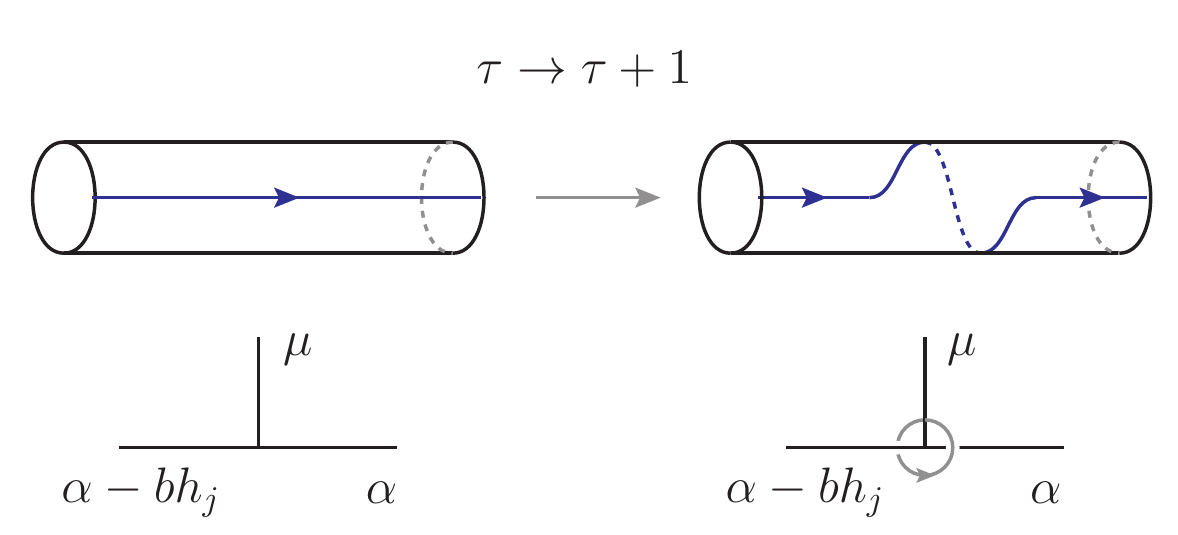}
\caption{\textit{Acting with the modular transformation $\tau \to \tau+1$ causes the defect line to wrap once around the tube. The corresponding braiding operation is shown in the second line.}}
\label{fig:wrapping_tube}
\end{figure}

Now consider a modification of this operator as shown in figure~\ref{fig:wrapping_tube}. Transporting the chiral vertex operator once around the tube introduces an intermediate phase into the operator given by the braiding factor
\be
e^{2 \pi i \left( \Delta(\al-bh_j) - \Delta(\al)-\Delta(\mu) \right)} =  (-\q)^{N-1}\af_j\, .
\ee 
In what follows, we introduce one additional framing factor $(-1)^{N-1} \q^{1/N-N}$ for each such rotation, so that the contribution becomes $\q^{1/N-1}\af_j $. The inclusion of this framing factor is suggested by consistency with the monodromy of the conformal blocks under the modular transformation $\tau\to\tau+1$, which is given by $e^{2 \pi i \Delta(\al)}$. Conjugating by the monodromy, we find
\be
e^{-2\pi i \Delta(\al)} \, \Delta_j \, e^{2\pi i \Delta(\al)} = \q^{1/N-1}\af_j \, \Delta_j 
\ee
in agreement with our convention for the framing. In summary, we have
\be
\begin{alignedat}{2}
 \mathfig{0.27}{defects/dyon_loop_pp}  &=  \;   \sum_{j=1}^N \Big[ \q^{1/N-1}\af_j \Big] \, c_j \Delta_j \\
 \mathfig{0.27}{defects/dyon_loop_mp} &=  \;  \sum_{j=1}^N \Big[ \q^{1-1/N}\af^{-1}_j \Big] \,  c_j\Delta_j  \,.
\end{alignedat}
\ee
In order to obtain the Verlinde operator for the orientation reversed defect line, one should replace $\af_j \to \af_j^{-1}$.

\begin{figure}[h]
\centering
\includegraphics[height=4cm]{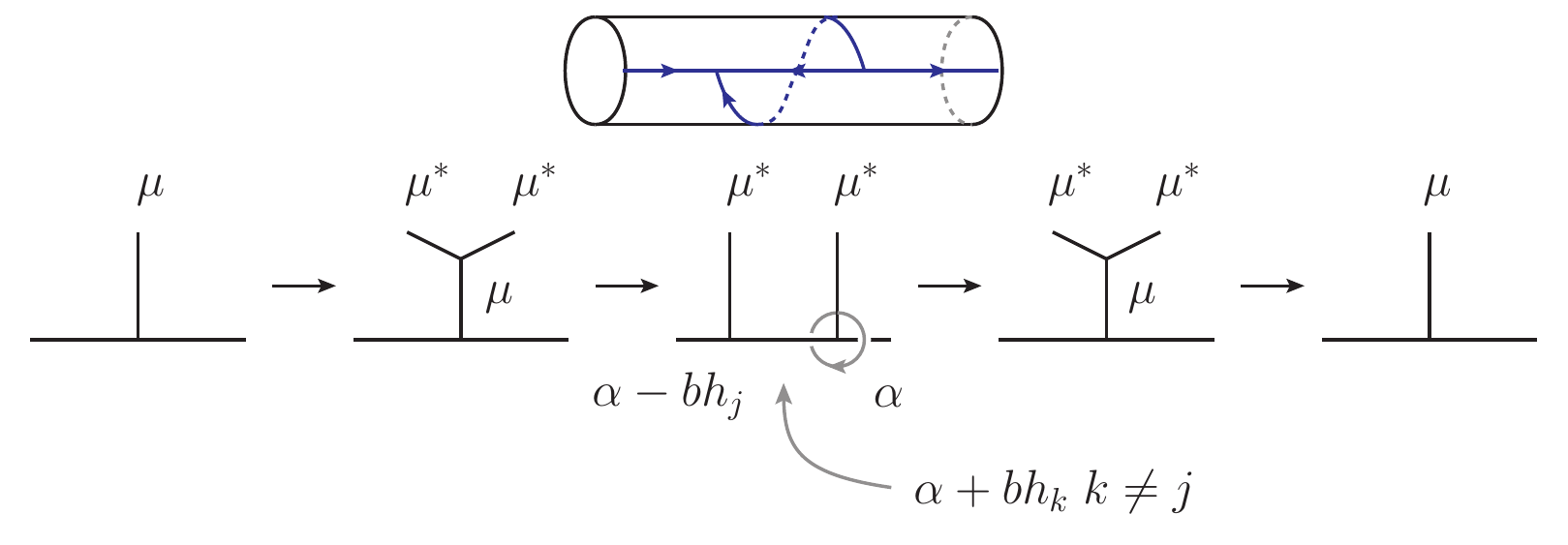}
\caption{\textit{Sequence of fusion and braiding operations for a topological defect network in $A_2$ Toda theory on a cylinder.}}
\label{fig:wrapping_tube}
\end{figure}

Let us now specialise to the case of $A_2$ and consider the sequence of operations shown in figure, which corresponds to a bubble network that surrounds the cylinder. Consider the term where the momentum on the left is $\al-bh_1$, that is the term multiplying the shift operator $\Delta_1$. The intermediate s-channel momenta are then constrained by the operator product expansion to be $\al+bh_k$ where $k=2,3$. In this intermediate step there is a braiding phase 
\bea
e^{2\pi i \left(\Delta(\al+bh_k)-\Delta(\al)-\Delta(\mu^*) \right)}= e^{-2\pi b a_k + 2 \pi i b^2}\, .
\eea
Now, combining the relevant fusion matrices and multiplying by the expectation value of a contractible bubble, we find
\bea
e^{2 \pi i b^2} \left[ e^{-2\pi b a_1} \frac{\sin\pi(iba_{12}+qb)}{\sin\pi (iba_{12})} + e^{-2\pi b a_2} \frac{\sin\pi(iba_{21}+qb)}{\sin\pi (iba_{21})}\right] 
= -  \q^3 (\af_2+\af_3)\, .
\eea
Braided instead the chiral primary on the left instead, we the answer $-\q^{-7/3}(\af_2+\af_3)$, which differs from the above by two powers of the framing factor $-\q^{8/3}$ for $A_2$. Here we are going to choose the intermediate framing, where we have $\q^{1/3}(\af_2+\af_3)$. The remaining cases are obtained by cyclic permutation of $\{1,2,3\}$. Once again, we note that braiding in the opposite direction inverts the whole factor, so that in summary
\be
\begin{alignedat}{2}
 \mathfig{0.27}{defects/dyon_web_pp}  &=  \;  \sum_{j=1}^3 \Big[ \q^{1/3}  \sum_{i\neq j} \af_i   \Big] \, c_j \Delta_j  \\
 \mathfig{0.27}{defects/dyon_web_mp} &=  \; \sum_{j=1}^3  \Big[ \q^{-1/3} \sum_{i\neq j} \af^{-1}_i \Big] \, c_j \Delta_j   \, .
\end{alignedat}
\ee
and reversing the orientation of the network sends $\af_j \to \af_j^{-1}$. 

\medskip

Given the Verlinde operators computed above for networks on a cylinder, we can now derive generalised skein relations for computing the composition of two Verlinde operators. Graphically, we express the composition $\cO_1 \cdot \cO_2$ by superimposing the network corresponding to $\cO_1$ on top of the network corresponding to $\cO_2$. 

Let us first recall what happens in Liouville theory. In this case, we set $N=2$ and identify $\af_1 = \af_2^{-1} = \af$ with $\Delta: \af \to \q^{-1} \af$. Now, composing the Verlinde operators corresponding to defect lines wrapping around and passing along the cylinder, we find
\be
\mathfig{0.25}{defects/comp_abelian_1} = \q^{-1/2}  \mathfig{0.25}{defects/dyon_loop_abelian_1}  + \q^{1/2} \mathfig{0.25}{defects/dyon_loop_abelian_2}  
\ee
while reversing the order of composition interchanges $\q\to\q^{-1}$ in the above equation. The above equations can be expressed as a rule for resolving a crossing
\be
\mathfig{0.09}{defects/skein_abelian_2} = \, \q^{-1/2} \mathfig{0.09}{defects/skein_abelian_3} + \, \q^{1/2} \mathfig{0.09}{defects/skein_abelian_4}
\ee
found in~\cite{Drukker:2009id}. The composition of any two Verlinde operators can be expanded superimposing the closed loops, resolving each crossing according the skein relation, and then removing any closed contractible loops according to~\eqref{eq:closedcontloop}.

Now consider $A_{2}$ Toda theory. Performing the same composition of defect lines around and along the cylinder, we now find
\bea
\mathfig{0.25}{defects/comp_nonabelian_1} = \q^{-2/3}  \mathfig{0.25}{defects/dyon_loop_pp}  + \q^{1/3} \mathfig{0.25}{defects/dyon_web_pp} \\ 
\mathfig{0.25}{defects/comp_nonabelian_2} = \q^{2/3}  \mathfig{0.25}{defects/dyon_loop_mp}  + \q^{-1/3} \mathfig{0.25}{defects/dyon_web_mp} 
\eea
This result can again be expressed as a generalised skein relation for resolving the crossing. This time, resolving the crossing necessarily introduces networks with junctions,
\be
\mathfig{0.09}{defects/skein_2} = \, \q^{-2/3} \mathfig{0.09}{defects/skein_3} + \, \q^{1/3}  \mathfig{0.09}{defects/skein_4}
\label{su3skein1}
\ee
\be
\mathfig{0.09}{defects/skein_1} = \, \q^{2/3} \mathfig{0.09}{defects/skein_3} + \, \q^{-1/3} \mathfig{0.09}{defects/skein_4}
\label{su3skein2}
\ee
Any composition of topological defects can be computed by superimposing the networks and resolving each crossing using the above generalised skein relations. The result may contain contractible networks that can be removed according to the rules derived in subsection~. The result is an expansion in irreducible defects.  

\begin{figure}
\centering
\includegraphics[height=2.5cm]{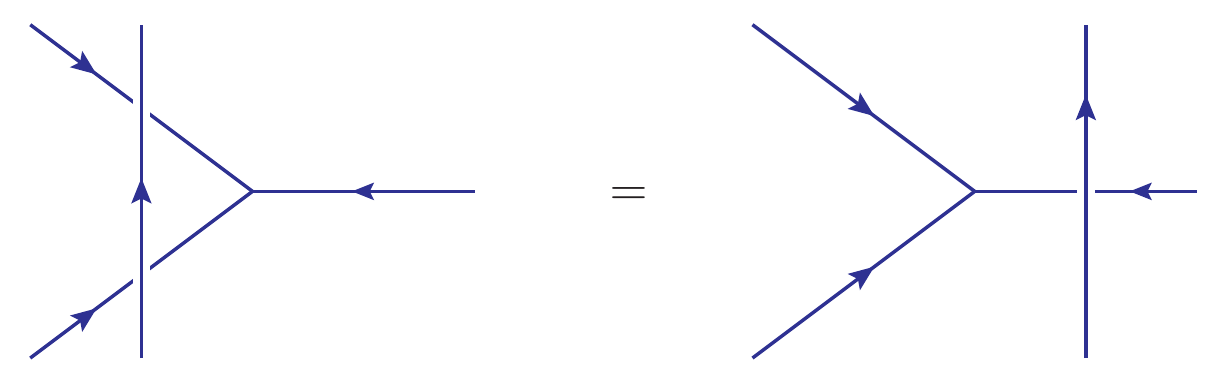}
\caption{\emph{Example of topological nature of the defects.}}
\label{yang_baxter}
\end{figure}

For $A_2$ Toda theory, the rules for removing contractible networks and the generalised skein relations together generate the `spider' encoding the representation theory of the quantum group $U_{\q}(\mathrm{sl}(3))$~\cite{Kuperberg:kx}.  An important consistence check of the proposal here is that these relations imply Verlinde operators depend only on the homotopy classes of the networks, that is, Verlinde operators introduce topological defects. An example of such a relation is shown in figure~\ref{yang_baxter}. The same relations ensure the topological invariance of quantum link invariants.

\section{Torus with puncture /  $\cN=2^*$ Theory}
\label{Section:N=2*}

In this section, we consider the relationship between topological defect networks and the expectation value of supersymmetric loop operators in $\cN=2$ gauge theories on an ellipsoid. We focus on a simple and concrete example: the $\cN=2^*$ theory consisting of an $\cN=2$ vectormultiplet and a massive hypermultiplet in the adjoint representation. The partition function of this theory on an ellipsoid is captured by a correlation function of a single semi-degenerate primary field in Liouville / Toda conformal field theory on a torus with puncture.

\subsection{Toda Correlator / Four-Sphere Partition Function}

Consider the correlation function of a semi-degenerate primary field $V_{\nu}$ in $A_{N-1}$ Toda theory on a torus with complex structure $\tau$. We take the semi-degenerate momentum to be
\be
\nu = N \left( \frac{q}{2} + i m \right) \omega_{N-1}
\label{mudef}
\ee
where $m \in \mathbb{R}_{\geq 0}$ is a real parameter. 

\begin{figure}[h]
\centering
\includegraphics[height=2cm]{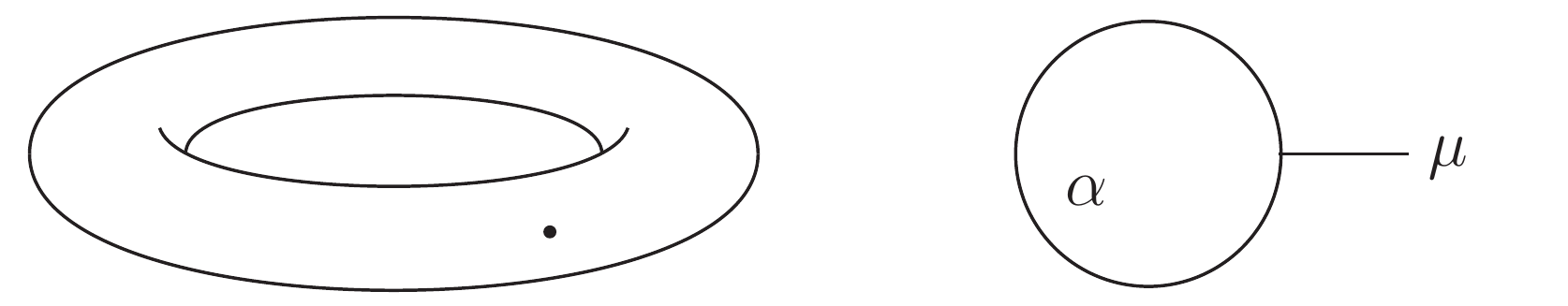}
\caption{\emph{The four-sphere partition function of $\cN=2^*$ theory is obtained by a correlation function in Toda CFT on a torus with insertion of a semi-degenerate vertex operator. The correlator is computed by choosing the pants decomposition shown on the right.}}
\label{fig:pantsdecomp}
\end{figure}

In order compute this correlator, we pick a pants decomposition with generic non-degenerate momentum $\al = Q+ia$ in the intermediate channel, as shown in figure~\ref{fig:pantsdecomp}. The correlation function is expanded
\be
\la V_{\nu} \ra_\tau =  \int da \, C(\al,\bar \al,\nu) \, \bar\cF_{\al,\nu}(\tau) \, \cF_{\al,\nu}(\tau)
\ee
where $C(\al,\bar \al, \nu)$ is the three-point function and $\cF_{\al,\nu}(\tau)$ are the $W_N$-algebra conformal blocks for our choice of pants decomposition. The contour for the integral is the Cartan subalgebra, $a \in \mathbb{R}^{N-1}$. Expanding in powers of $e^{2\pi i \tau}$, the conformal blocks are normalised so that the leading term is one.

The generic three-point function in Toda theory is unknown for $N>2$. However, the three-point function involving one semi-degenerate momentum of the form $\nu=\kappa\, \omega_{N-1}$ has been computed in~\cite{Fateev:2005gs,Fateev:2007ab} by the bootstrap method. For $N=2$ this result reproduces the generic Liouville three-point function~\cite{Dorn:1994xn,Zamolodchikov:1995aa}. The three-point function $C(\al,\bar\al,\nu)$ with the choice of semi-degenerate momentum~\eqref{mudef} can be expressed in a factored form 
\be
C(\al,\bar\al,\mu) = \mu (a) \, \left| \, \frac{ \prod\limits_{i<j}^N \Gamma_b\left(\frac{q}{2}+ia_{ij}+im \right) }{ \prod\limits_{i<j}^N \Gamma_b\left(q+ia_{ij}\right)  \Gamma_b(q-ia_{ij}) } \, \right|^2
\ee
up to a certain prefactor that is independent of the internal momentum $\al$ that can be absorbed in the primary field $V_{\nu}$. In the above expression we have pulled out a new integration measure
\be
d \mu (a)  = da \prod_{i<j} 2 \sinh\left( \pi b a_{ij} \right) \, 2\sinh\left( \pi b^{-1} a_{ij} \right)\, .
\ee
The notation $\Gamma_b(x)$ denotes the double gamma function, some of whose properties are summarised in appendix~\ref{appendix:special}. 

Given this factorisation of the three-point function, it is convenient to define new renormalised $W_N$-algebra conformal blocks
\be
\cG_{a,m}(\tau) = \frac{ \prod\limits_{i<j}^N \Gamma_b\left(\frac{q}{2}+ia_{ij}+im \right) }{ \prod\limits_{i<j}^N \Gamma_b\left(q+ia_{ij}\right)  \Gamma_b(q-ia_{ij}) }\cF_{\al,\mu}(\tau)
\label{eq:normdef}
\ee 
so that the correlator is
\be
\int da \, \mu(a) \, \bar\cG_{a,m}(\tau) \cG_{a,m}(\tau)\, .
\ee
This choice of normalisation coincides with a common one made in Liouville theory when $N=2$ (see for example~\cite{Drukker:2009id}) and is designed to simplify the expressions for Verlinde operators so that they constructed only from ratios of trigonometric functions of the parameters $a$ and $m$.

This correlator captures the ellipsoid partition function of $\cN=2^*$ theory with gauge algebra $\mathrm{su}(N)$. This was computed independently by supersymmetric localisation in the case of a round four-sphere in the seminal work~\cite{Pestun:2007rz} (an important observation regarding the hypermultiplet mass in the localisation formula was also made in~\cite{Okuda:2010ke}) and was later extended to the ellipsoid in~\cite{Hama:2012bg}. The correspondence between the parameters are summarised below:
\begin{enumerate}
\item The dimensionless coupling $b$ corresponds to the shape of the ellipsoid, defined by the embedding
\be
x_0^2 + \frac{1}{b^2}(x_1^2+x_2^2) + b^2 ( x_3^2+x_4^2 ) = 0
\ee
in $\mathbb{R}^5$ with euclidean coordinates $x_0,x_1,\ldots,x_4$.
\item The complex structure of the torus corresponds to the UV holomorphic gauge coupling,
\be
\tau = \frac{4\pi i}{g^2}+ \frac{\theta}{2\pi} \, .
\ee
\item The parameter $m$ is the mass of the adjoint hypermultiplet.
\item A choice of pants decomposition corresponds to a duality frame in $\cN=2^*$ theory and the mapping class group $SL(2,\mathbb{Z})$ corresponds to the duality group. 
\item The integration over the internal momenta $\al = Q+ia$ becomes an integral over the Coulomb branch, parametrised by the vacuum expectation value $\phi = \bar\phi = -\frac{i}{2}a$ of the vectormultiplet scalar.
\end{enumerate}

We expect that our choice factorisation used to define the renormalised $W_N$-algebra conformal block $\cG_{a,m}(\tau)$ has a simple interpretation: it corresponds to the partition function on half of the four-sphere $\{x_0 >0\}$ with Dirichlet boundary conditions for the vectormultiplet. This boundary condition is specified by the expectation value $a$ of the vectormultiplet scalar. Similarly, the conjugate renormalised block $\bar\cG_{a,m}(\tau)$ corresponds to the partition function on $\{x_0<0\}$. The complete partition function is reconstructed by introducing a three-dimensional $\cN=2$ vectormultiplet on the boundary $\{x_0=0\}$, whose partition function is exactly the measure $\mu(a)$, and integrating over the boundary condition $a$. 

The standard decomposition into components of the Nekrasov partition function is discussed in appendix~\ref{appendix:special}.

\subsection{Supersymmetric Loop Operators}

In the first half of the paper, we considered a class of Verlinde operators acting on the space of $W_N$-algebra conformal blocks by shifting the internal momentum. These operators are expected to compute the expectation value of supersymmetric loop operators in the four-dimensional gauge theory supported on the circle
\bea
 x_1 &= b \cos \varphi \\
 x_2 &= b \sin \varphi \\
 x_0 &= x_3 = x_4 = 0
\eea
where $0<\varphi<2\pi$~\footnote{In subsection~(\ref{sub:hopf}) we will also consider supersymmetric loop operators supported on the circle given by exchanging $b \leftrightarrow b^{-1}$ and the coordinate planes $\{ x_1,x_2\}$ and $\{x_3,x_4\}$. Thet are computed by Verlinde operators constructed by transporting chiral vertex operators with momentum of the form $-b^{-1}\lambda$ where $\lambda$ is a fundamental weight.}.

Focussing on a torus with simple puncture, we can always conjugate a Verlinde operator by the proportionality factor in equation~\eqref{eq:normdef} to obtain a difference operator acting on the renormalised conformal block $\cG_{a,m}(\tau)$. The correlation function in the presence of the topological defect corresponding to a Verlinde operator $\cO$ then has the generic form
\be
 \int d \mu(a) \,  \bar\cG_{a,m}(\tau) \, \left(  \cO \cdot \cG_{a,m}(\tau)  \right) 
\ee
where $\cO$ is a difference operator acting on the internal momentum $\al = Q + ia$. The operators $\cO$ are self-adjoint with respect to the measure $d\mu(a)$. In all known examples, the expectation values of supersymmetric loop operators can also be expressed in this form. For example, Wilson loops correspond to multiplicative operators~\cite{Pestun:2007rz}, while 't Hooft loops are genuine finite difference operators~\cite{Gomis:2010kv}.

The Verlinde operators have natural operations of addition and composition. The algebra generated by these operations corresponds to the operator product expansion of supersymmetric loop operators in the four-dimensional gauge theory. Note that the support of the loop operator can be deformed supersymmetrically away from the boundary,
\bea
x_0 &= \cos \rho \\
x_1 &= \sin \rho \cos \varphi \\
x_2 &= \sin \rho \sin \varphi \\
x_3 &= x_4 =0
\eea
with $0 < \rho < \pi$ varying between the north and sound poles. Taking the limit $\rho_1 ,\rho_2 \to \frac{\pi}{2}$ as two loop operators $\cO_1$ and $\cO_2$ coincide at the equator, the expectation depends only on the order in which they are brought to the equator. This is the ordering of the operators in the composition $\cO_1 \circ \cO_2$. In the previous section, we saw that any such composition can be expanded as a linear combination of irreducible Verlinde operators $\sum_j c_j \, \cO_j$ by resolving crossings according to set of generalised Skein relations. Once the dictionary between topological defects and loop operators is known, this provides a way to compute of the operator product expansion.

Before mapping out the dictionary, let us recall the classification of supersymmetric loop operators~\cite{Kapustin:2005py}. Setting aside for the moment the global structure of the gauge group and mutual locality constraints~\footnote{The mutual locality constraints arise in the present context when considering supersymmetric loop operators supported on Hopf linked circles in the equator. This is discussed below in subsection~\ref{sub:hopf}.}, the supersymmetric loop operators are labelled by a pair of weights $(\lambda_e,\lambda_m)$ such that pairs related by Weyl transformations are equivalent
\be
(\lambda_e,\lambda_m) \sim (w(\lambda_e),w(\lambda_m)) \qquad w \in S_N \, .
\ee  
It is sometimes convenient to repackage this information by transforming the magnetic weight $\lambda_m$ to a dominant integral weight by a Weyl transformation. The remaining Weyl transformations leaving $\lambda_m$ invariant are those of the commutant of $\lambda_m$ in $A_{N-1}$, which is the unbroken gauge algebra on the support of the operator. They can be used to transform $\lambda_e$ to a dominant integral weight of the unbroken gauge algebra. In summary, we can specify an irreducible representation of $A_{N-1}$ with highest weight $\lambda_m$ and an irreducible representation of the commutant of $\lambda_m$ in $A_{N-1}$ with highest weight $\lambda_e$.

The expectation value of the supersymmetric loop operator labelled by $(\lambda_e,\lambda_m)$ is defined by the path integral with a boundary condition in the vicinity of the loop given by (introducing local transverse coordinates $\{x^1,x^2,x^3\}$ to $S^1_{\varphi}$)
\begin{alignat}{2}
F_{ij} &= \frac{\lambda_m}{2} \epsilon_{ijk} \frac{x^k}{|\, \vec x\, |^3 } &\qquad  \phi - \bar\phi &= \frac{\lambda_m}{2 | \, \vec x\,  |}  \\
F_{\varphi i} &= \frac{i g^2}{8\pi} \lambda_e \, \frac{x^i}{|\, \vec x\, |^3}  &\phi + \bar\phi &= \frac{ig^2}{8\pi}\lambda_e\frac{1}{|\, \vec x \, |}\, .
\label{dyonbc}
\end{alignat}
together with an insertion
\be
\tr_{\lb_e} \, \mathrm{P} \, \exp i \oint_{S^1_{\varphi}} d\varphi \left( A_\varphi - b(\phi+\bar\phi) \right) 
\ee
where the trace is taken in the representation of the unbroken gauge algebra specified by the highest weight $\lambda_e$. Let us emphasize that the data $(\lambda_e,\lambda_m)$ are defined only up to Weyl transformations, which are realised by gauge transformations of the boundary condition~\eqref{dyonbc}. This boundary condition is correct when $\mathrm{Re}(\tau) = 0$. In the presence of non-zero theta-angle one should replace $\lb_e \to \lb_e + (\theta / 2\pi) \lb_m$ according to the Witten effect.

Finally, the supersymmetric loop operator labels $(\lb_e,\lb_m)$ transform in the fundamental representation of the duality group $SL(2,\mathbb{Z})$~\cite{Kapustin:2005py}. The duality group is generated by the transformations $S:\tau\to-1/\tau$ and $T:\tau\to\tau+1$ with
\bea
S &: (\lb_e,\lb_m) \mapsto (\lb_m,-\lb_e) \\
T &: (\lb_e,\lb_m) \mapsto (\lb_e,\lb_m+\lb_e) \, .
\label{duality}
\eea
They obey the relations $S^2 = C$ and $(ST)^3=C$ where the conjugation is defined by $C:(\lb_e,\lb_m) \to (-\lb_e,-\lb_m)$. The conjugation can be undone by a Weyl transformation and therefore acts trivially on the loop operators. To realise these transformations exactly for supersymmetric loop operators on a four-sphere requires a slightly unconventional choice of basis in the space of Wilson loops, as explained below.


\subsection{Wilson Loops}

Supersymmetric Wilson loops correspond to vanishing magnetic weight $\lambda_m=0$. They are thus classified by irreducible representations of $A_{N-1}$ with highest weight $\lambda_e$. The expectation value of a supersymmetric Wilson loop on the ellipsoid has been computed in~\cite{Hama:2012bg}, generalising the computation on a round four-sphere~\cite{Pestun:2007rz}. 

The result of the computation~\cite{Hama:2012bg} is to evaluate the supersymmetric Wilson loop on the expectation value $\phi = \bar\phi = -\frac{1}{2} a$ and insert the answer into the matrix integral. This leads to the factor $\tr_{\lb_e} \left( e^{-2\pi b a} \right)$ where the trace is taken in the irreducible representation with highest weight $\lb_e$. For example, for the $r$-th fundamental representation with highest weight $\lb_e = \omega_r$, we insert 
\be
\tr_{\omega_r}\left( e^{-2\pi b a} \right)
 = \sum_{j_1 < \ldots < j_r} e^{-2\pi b (a_{j_1}+\cdots+a_{j_r})}\, .
\ee
In other words, this is simply a multiplicative operator acting on the renormalised conformal blocks $\cG_{a,m}(\tau)$.

\begin{figure}[htp]
\centering
\includegraphics[height=2.5cm]{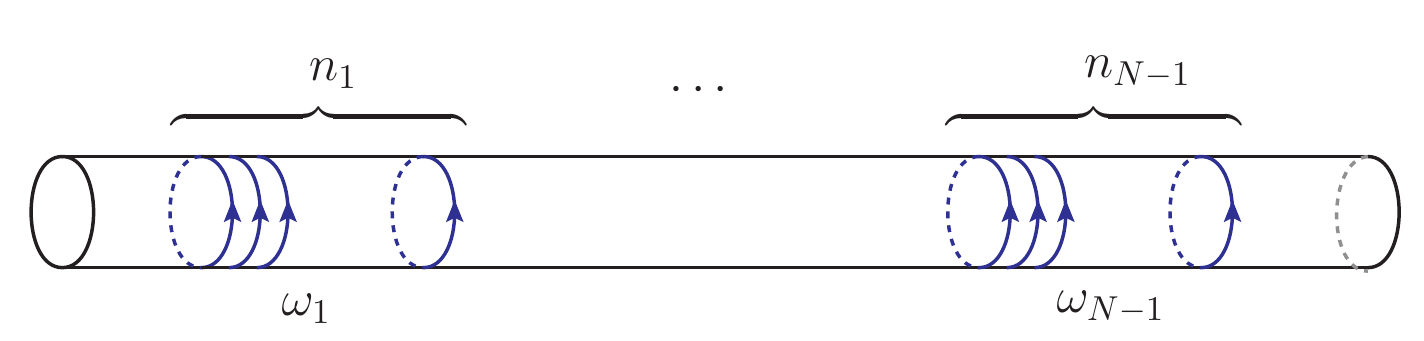}
\caption{\emph{Supersymmetric Wilson loops are in one-to-one correspondence with parallel irreducible defects wrapping the tube of the pants decomposition. Here and in what follows, we abuse graphical notation by cutting the torus according to the pants decomposition and drawing it as a cylinder.}}
\label{fig:genericwl}
\end{figure}

The Verlinde operators corresponding to supersymmetric Wilson loops in $\cN=2^*$ were studied in Liouville theory in~\cite{Drukker:2009id,Alday:2009fs} and Toda theory references in~\cite{Gomis:2010kv,Passerini:2010pr}. For constructing the full dictionary with irreducible Verlinde operators, we find it convenient to work with an alternative basis of supersymmetric Wilson loops: for a generic dominant integral weight $\lambda_e = \sum_{j} n_j \omega_j$ we insert $n_j$ powers of the supersymmetric Wilson loop in the $j$-th fundamental representation,
\be
\prod_{j=1}^{N-1} \Bigg[ \, \sum_{j_1 < \ldots < j_r} e^{-2\pi b (a_{j_1}+\cdots+a_{j_r})} \, \Bigg]^{n_j} \, .
\label{wldef}
\ee
This corresponds to an invertible linear transformation on the algebra of supersymmetric Wilson loops. 

The corresponding Verlinde operator is then constructed from $n_j$ parallel defect lines labelled by $j$ around the tube of the pants decomposition, as shown in figure~\ref{fig:genericwl}. Supersymmetric Wilson loops in the standard basis are obtained by fusing all of the parallel defects and extracting unique copy of the representation with highest weight $\lb_e$. This would be obtained by transporting a single primary field with momentum $\mu = -b\lb_e$ along the tube of the pants decomposition.


\subsection{'t Hooft Loops}

Supersymmetric 't Hooft loops are characterised by $\lb_e=0$ and so are labelled by a dominant integral magnetic weight $\lb_m$. Since Wilson loops and 't Hooft loops are related by $S:(\lb_e,\lb_m) \to (\lb_m,-\lb_e)$, we expect the Verlinde operator for the generic magnetic weight $\lb_m = \sum_j n_j \omega_j$ to be constructed from $n_j$ parallel defect lines labelled by $j$ passing along the tube of the pants decomposition.

\begin{figure}[htp]
\centering
\includegraphics[height=2.8cm]{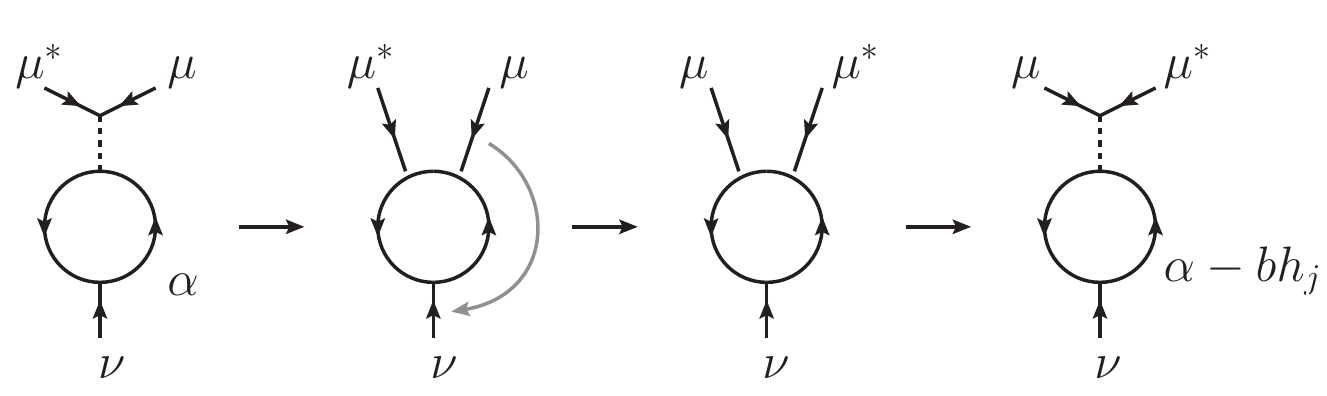}
\caption{\emph{The sequence of fusion and braiding operations required to compute the Verlinde operator corresponding to the 't Hooft loop of minimal charge in the $\cN=2^*$ theory.}}
\label{thooftsequence}
\end{figure}

Let us concentrate on the 't Hooft loop with minimal magnetic weight, $\lambda_m = \omega_1$. The corresponding Verlinde operator was identified in Liouville theory in~\cite{Drukker:2009id,Alday:2009fs} and extended to $A_{N-1}$ Toda theory in~\cite{Gomis:2010kv} and corresponds to a defect line running along the length of the pants decomposition. It is computed by the sequence of fusion and braiding moves shown in figure~\ref{thooftsequence}. 

In particular, the result obtained in~\cite{Gomis:2010kv} is a difference operator acting on the standard unnormalised $W_N$-algebra conformal blocks $\cF_{a,m}(\tau)$. To find an operator acting on the normalised conformal blocks $\cG_{a,m}(\tau)$, we must conjugate by the normalisation factor in equation~\eqref{eq:normdef}. A summary of this computation is presented in appendix~\ref{appendix:special}~\footnote{The result of this computation was obtained in the course of another project in collaboration with Martin Fluder, Lotte Hollands and Paul Richmond.}. The result is that the difference operator labelled by $(0,\omega_1)$ now involves only ratios of trigonometric functions,
\bea
\mathfig{0.27}{defects/thooft_loop_p}  &= \sum_{j=1}^n \prod_{k\neq j} \frac{\sin\pi b( \frac{q}{2}+ia_{kj}-im) }{ \sin \pi b( i a_{kj} )} \Delta_j \\
& = \sum_{j=1}^n \prod_{k \neq j} \frac{\tf\,\af_k -  \tf^{-1} \af_j}{\af_k-\af_j} \, \Delta_j
\label{eq:fundamentalthooft1}
\eea
where in the second line we have introduced for convenience a parameter $\tf = e^{-\pi b( m +iq/2)}$ encoding the hypermultiplet mass deformation, along with the parameters $\af_j = e^{-2\pi b a_j}$. The anti-fundamental 't Hooft loop is obtained by transporting the same primary in the opposite direction and the result is given by replacing $\af_j \to \af_j^{-1}$ (this includes $\Delta_i \to \Delta_j^{-1}$) in the above. 

Although the required fusion and braiding matrices have not appeared in the literature, we expect that transporting the primary with momentum $\mu=-b\omega_r$ around the same path leads to the operator
\bea
\mathfig{0.27}{defects/thooft_loop_r}  & = \sum_{j_1 < \ldots < j_r} \prod_{\substack{ j \in \{j_1,\ldots,j_r\} \\ k \neq \{ j_1,\ldots,j_r\} }} \frac{\tf\,\af_k -  \tf^{-1} \af_j}{\af_k-\af_j} \, \Delta_{j_1} \ldots \Delta_{j_r} 
\label{eq:fundamentalthooft2}
\eea
and corresponds to the 't Hooft loop in the $r$-th anti-symmetric tensor representation, labelled by $(0,\omega_r)$. For these representations, there are no monopole bubbling effects in the localisation computation and it is straightforward to show that this expression agrees with the localisation computation of~\cite{Gomis:2011pf} in the case of a round four-sphere, $b\to1$. This is demonstrated in appendix~\ref{appendix:special}.

For non-fundamental magnetic weight, there are monopole bubbling effects and the difference operators become more complicated. For example, let us consider the 't Hooft loop with magnetic weight $2\omega_1$. Reference~\cite{Gomis:2011pf} found in the case $N=2$ that the correct difference operator is found from the composition $\cO_{0,\omega_1} \cdot \cO_{0,\omega_1}$ of two copies of the fundamental 't Hooft loop operator, corresponding to two parallel defect lines. Extrapolating this result without proof to the case $N>2$, we claim that the difference operator is given by
\vspace{1mm}
\bea
& \mathfig{0.27}{defects/thooft_loop_double} = \sum_{j=1}^N \; \left[ \prod_{k \neq j}^N \frac{\tf\,\af_k -  \tf^{-1} \af_j}{\af_k-\af_j} \frac{\q^2\tf\,\af_k -  \tf^{-1} \af_j}{\q^2\af_k-\af_j}\right] \Delta_j^2 \\
& + \sum_{i<j}^N \left[ \prod_{k\neq i,j} \frac{\tf\,\af_k -  \tf^{-1} \af_i}{\af_k-\af_i}\frac{\tf\,\af_k -  \tf^{-1} \af_j}{\af_k-\af_j} \right]\, \left[ \frac{\tf\,\af_i -  \tf^{-1} \af_j}{\af_i-\af_j} \frac{\tf\,\af_j -  \q^2\tf^{-1} \af_i}{\af_j-\q^2\af_i}  +  i \leftrightarrow j \right] \Delta_i \Delta_j \, .
\label{symten}
\eea

\noindent Note that the weights of the symmetric tensor representation decompose into two Weyl orbits $\{2h_j,j=1,\ldots,N\}$ and $\{h_i+h_j, i<j\}$ appearing in the first and second lines respectively in equation~\eqref{symten}. Monopole bubbling is seen in the latter by the presence of a sum of terms multiplying each shift. 

Assuming that the correct difference operator for the 't Hooft loop with magnetic weight $2\omega_1$ is always given by the composition $\cO_{0,\omega_1} \cdot \cO_{0,\omega_1}$ rather than $(\cO_{0,\omega_1} \cdot \cO_{0,\omega_1}-1)$, this operator is related by S-duality to the product of two Wilson loops in the fundamental representation rather than a Wilson loop defined by a trace in the symmetric tensor representation. As explained in~\cite{Gomis:2011pf}, the origin lies in the natural resolution of the Bogomolnyi moduli space used in the localisation computation. This is the basic reason why we think it is more natural to use the basis of Wilson loops defined in equation~\eqref{wldef}.


\subsection{Dyonic Loops}

Let us now consider dyonic loop operators labelled by non-zero electric and magnetic weights $(\lambda_e,\lambda_m)$. For simplicity, we would like to avoid monopole bubbling effects and so focus on dyonic loops obtained by adding supersymmetric Wilson loops in the background of 't Hooft loop with magnetic charge $\lb_m=\omega_1$. 

In this case, the monopole singularity is proportional to
\be
\omega_1 = i \mathrm{diag} \left( 1-\frac{1}{N} , -\frac{1}{N} , \ldots, -\frac{1}{N} \right)
\ee
breaking the gauge algebra
\be
\mathrm{su}(N) \to \mathrm{s}( \mathrm{u}(1) \oplus \mathrm{u}(N-1))
\ee
on the support of the loop operator. The remaining Weyl transformations leaving the magnetic weight $\omega_1$ invariant can be used to transform the electric weight into the following form
\be
\lambda_e = m_1 \omega_1 -  \sum_{j=2}^{n-1} m_j \omega_j
\qquad 
m_1\in \mathbb{Z} \qquad  m_2,\ldots,m_N \in \mathbb{Z}_{\geq0} \, .
\ee
The first integer corresponds to adding an abelian Wilson loop of charge $m_1$ in the $\mathrm{u}(1)$ factor, while the positive integers $(m_2,\ldots,m_N)$ are Dynkin labels for an irreducible representation of the $\mathrm{u}(N-1)$ factor. In the same manner as for plain Wilson loops, we expect it is most natural choose a basis where the label $(m_2,\ldots,m_N)$ corresponds to inserting Wilson loops in the $(j-1)$-th anti-symmetric representations of $\mathrm{u}(N-1)$ raised to the $m_j$-th power.

\subsubsection{Easy Dyons}

The simplest dyonic operators are those with $(m_2,\ldots,m_N)=0$ corresponding to Wilson loops in the unbroken $\mathrm{u}(1)$. Such operators are related to the fundamental 't Hooft loop by some number of duality transformations, $\tau \to \tau+1$. As shown in section, we can compute these operators by conjugating with the monodromy $e^{2\pi i \Delta(\al)}$ of the conformal blocks,
\be
e^{-2\pi i \Delta(\al) } \, \Delta_j \, e^{2 \pi i \Delta(\al)} = \q^{1/N-1} \af_j \, .
\ee
Thus, the dyonic loop operator labelled by $(n\, \omega_1,\omega_1)$ is
\vspace{0.1cm}
\bea
\mathfig{0.32}{defects/dyon_loop_multiple} & = \sum_{j=1}^N \prod_{k \neq j}^N \frac{\tf\,\af_k -  \tf^{-1} \af_j}{\af_k-\af_j}  \,\big[ \q^{1/N-1} \af_j \big]^n \, \Delta_j \, .
\eea

This result can be compared with an exact localisation calculation in the limit $b\to1$. There is one term in the 't Hooft loop operator for each weight $\{h_j,j=1,\ldots,N\}$ in the fundamental representation, which are permuted by Weyl / gauge transformations. Let us denote the unbroken $\mathrm{u}(1)$ in the term corresponding to the weight $h_j$ by $\mathrm{u}(1)_j$. Then in the localisation computation, we must add the abelian Wilson loop
\be
\tr_{U(1)_j} e^{2\pi n (ia - \frac{h_j}{2})} =  e^{i\pi n (1/N-1)} e^{-2\pi n a_j } \,. 
\ee 
in the $j$-th term. This is in agreement with the prediction of Toda in the limit $b\to1$.

\subsubsection{Hard Dyons}

The exhausts the spectrum of dyonic loop operators with minimal magnetic charge in the case $N=2$. However, for $N>2$ there exists many more dyonic loop operators labelled by non-zero $(m_2,\ldots,m_N)$. They all correspond to Verlinde operators constructed from networks with junctions. For the case of $N=3$, the gauge algebra is broken to $\mathrm{s}(\mathrm{u}(1) \oplus \mathrm{u}(2))$. In the frame where $\lb_m = \omega_1$, the remaining Weyl transformations consist of the identity $\mathbbm{1}$ and the reflection $\omega_1-\omega_2\leftrightarrow \omega_2$. Thus, starting from a generic pair $(m_1 \, \omega_1 - m_2 \, \omega_2,\omega_1)$ we can always ensure $m_2\geq0$ by such a transformation, as claimed above.

Let us first consider the case of the dyonic loop operator labelled by the weights $(-\omega_2,\omega_1)$. We claim that the correct Verlinde operator is given by
\bea
 \mathfig{0.27}{defects/dyon_web_pp}  & =    \sum_{j=1}^3 \, \prod_{k \neq j}^3 \frac{\tf\,\af_k -  \tf^{-1} \af_j}{\af_k-\af_j}   \, \Big[   \q^{1/3} \sum_{k\neq j}^3 \af_k   \Big] \, \Delta_j \\
\eea
Let us now compare with the corresponding localisation computation. From this perspective, we have an additional fundamental Wilson loop in the unbroken $\mathrm{u}(2)$ factor. Let us denote the unbroken $\mathrm{u}(2)$ in the Weyl frame where the monopole singularity is proportional to $h_j$ by $\mathrm{u}(2)_j$. Then in the $j$-th term of the operator, we must insert an additional factor
\be
\tr_{U(2)_j} e^{2\pi \left(ia -\frac{h_j}{2} \right)}  = e^{i \pi/3} \sum_{k\neq j}^3 e^{-2\pi a_k}\, .
\ee
This agrees with our proposal in the limit $b \to 1$. 

It is now straightforward to obtain any dyonic loop operator related to this by $T$ transformations. For example, the dyonic loop operator $(\omega_2,\omega_1)  \sim (\omega_1-\omega_2,\omega_1)$ is obtained from $(-\omega_2,\omega_1)$ by
\bea
 \mathfig{0.27}{defects/dyon_web_mp} & = \mathfig{0.27}{defects/dyon_web_mp_2} \\
 & =  \sum_{j=1}^3 \, \prod_{k \neq j}^3 \frac{\tf\,\af_k -  \tf^{-1} \af_j}{\af_k-\af_j} \, \Big[ \q^{-2/3} \af_j \Big]  \, \Big[   \q^{1/3} \sum_{k\neq j} \af_k   \Big] \, \Delta_j \\
 & = \sum_{j=1}^3 \, \prod_{k \neq j}^3 \frac{\tf\,\af_k -  \tf^{-1} \af_j}{\af_k-\af_j}  \, \Big[   \q^{-1/3} \sum_{k\neq j} \af^{-1}_k   \Big] \, \Delta_j 
 \eea
where it is now critical that $\af_1 \af_2 \af_3 = 1$. 

Extending this argument, it is now straightforward to identify the generic dyonic loop operator with minimal magnetic weight $(m_1\omega_1-m_2\omega_2,\omega_1)$ where $m_1\in \mathbb{Z}$ and $m_2\in \mathbb{Z}_{\geq0}$ with the following Verlinde operator
\bea
 & \mathfig{0.5}{defects/dyon_loop_generic} \\
 &  \;\;\;\;\;\;\;\;\;\;\;\;\;\;\;\;\;\; = \sum_{j=1}^3 \, \prod_{k \neq j}^3 \frac{\tf\,\af_k -  \tf^{-1} \af_j}{\af_k-\af_j} \, \Big[ \q^{-2/3} \af_j \Big]^{m_1}  \, \Big[   \q^{1/3} \sum_{k\neq j}^3 \af_k   \Big]^{m_2} \, \Delta_j\, .
 \label{dyonmin}
 \eea
This expression is in agreement with the appropriate additional factor in the localisation computation
\bea
\tr _{U(1)_j } e^{2\pi m_1 \left(ia- \frac{h_j}{2} \right)}\, \left[ \tr_{U(2)_j} e^{2 \pi \left( i a -\frac{h_j}{2} \right)} \right]^{m_2}  = \Big[ e^{-2i\pi/3} e^{-2\pi a_j} \Big]^{m_1}  \, \Big[   e^{i\pi/3} \sum_{k\neq j}^3 e^{-2\pi a_k}   \Big]^{m_2}
\eea
in the limit $b\to1$. 

This exhausts the spectrum of dyonic loop operators with minimal magnetic weight when $N=3$. There is an identical construction for dyonic loops with the conjugate magnetic weight $\omega_2$ by reversing the orientation of the networks. However, for dyons with non-miniscule magnetic weight, the four-dimensional physics involves monopole bubbling contributions and the operators become more complicated. Likewise, the spectrum of networks required is much richer. We leave the investigation of such operators for future work.


\subsection{Skein Relations  / Operator Product Expansion}

In section~(\ref{sub:Skein Relations}) we have derived a set of generalised skein relations to resolve the composition $\cO_1 \cdot \cO_2$ of two Verlinde operators, and we have argued above that this corresponds to an operator product expansion of the corresponding loop operators. Let us now substantiate this claim.

As above, we focus on the $\cN=2^*$ theory with gauge algebra $\mathrm{su}(3)$. Let us compose the Verlinde operators corresponding to fundamental Wilson loop and the 't Hooft loop operator with minimal magnetic weight. From the generalised skein relations~\eqref{su3skein1} and ~\eqref{su3skein2}, or directly from the operators above, we find
\bea
\mathfig{0.25}{defects/comp_nonabelian_1} = \q^{-2/3}  \mathfig{0.25}{defects/dyon_loop_pp}  + \q^{1/3} \mathfig{0.25}{defects/dyon_web_pp} \\
\mathfig{0.25}{defects/comp_nonabelian_3} = \q^{2/3}  \mathfig{0.25}{defects/dyon_loop_pp}  + \q^{-1/3} \mathfig{0.25}{defects/dyon_web_pp} \, .
\label{skeinN=2*}
\eea
Let us write the Verlinde operator corresponding to the dyonic loop $(\lb_e,\lb_m)$ as $\cO_{\lb_e,\lb_m}$. Then equation~\eqref{skeinN=2*} becomes
\bea
\cO_{0,\omega_1}  \, \cO_{\omega_1,0} &= \q^{-2/3} \cO_{\omega_1,\omega_1} + \q^{1/3}\cO_{-\omega_2,\omega_1}\\
\cO_{\omega_1,0} \, \cO_{0,\omega_1} &= \q^{2/3} \cO_{\omega_1,\omega_1} + \q^{-1/3}\cO_{-\omega_2,\omega_1} 
\eea
for the difference operators.

From the generalised skein relations, this result can be immediately extended to the composition of the fundamental Wilson loop operator with any dyonic loop operator of minimal magnetic weight,
\bea
\cO_{\lb,\omega_1} \, \cO_{\omega_1,0} &= \q^{-2/3} \cO_{\lb+\omega_1,\omega_1} + \q^{1/3}\cO_{\lb-\omega_2,\omega_1}\\
\cO_{\omega_1,0} \, \cO_{\lb,\omega_1} &= \q^{2/3} \cO_{\lb+\omega_1,\omega_1} + \q^{-1/3}\cO_{\lb-\omega_2,\omega_1} \, .
\eea
which is readily checked directly from the explicit form of the operators given in equation~\eqref{dyonmin}. Similar equations can be obtained from the skein relations for composing operators with the anti-fundamental Wilson loop.

By taking linear combinations of the above equations we can extract either of the terms on the righthand side. For example, we could introduce $\q$-commutators
\bea
\big[ \, \cO_1 \, , \, \cO_2 \, \big]_{1} & = \q^{1/3}\, \cO_1\, \cO_2  - q^{-1/3}\,\cO_2\,\cO_1 \\
\big[ \, \cO_1 \, , \, \cO_2 \, \big]_{2} &= \q^{-2/3}\, \cO_1\, \cO_2  - q^{2/3}\,\cO_2\,\cO_1
\eea
such that
\bea
\big[ \, \cO_{\omega_1,0} \, , \, \cO_{\lb,\omega_1} \, \big]_1 & = (\q-\q^{-1}) \cO_{\lb+\omega_1,\omega_1} \\
\big[ \, \cO_{\omega_1,0} \, , \, \cO_{\lb,\omega_1} \, \big]_2 & = -(\q-\q^{-1}) \cO_{\lb-\omega_2,\omega_1} \, .
\eea
In this way, the fundamental Wilson loop can be interpreted as a creation operator, raising the electric weight by $\omega_1$ and $-\omega_2$ with respect to the $\q$-commutators $[\; ,\, ]_1$ and $[\; ,\, ]_2$ respectively. Similarly, the anti-fundamental Wilson loop is an annihilation operator with respect to the same $\q$-commutators. 

The simplicity of the above operator product expansions came because there was only one crossing of the networks. For dyonic loop operators with non-miniscule magnetic weight, composing with a fundamental Wilson loop necessarily involves resolving multiple crossings. Thus, the operator product expansion is more complicated. Nevertheless, once the dictionary has been found, the generalised skein relations should provide a systematic way to compute the operator product expansion of supersymmetric loop operators. We hope to return to this question in future work.

\subsection{'t Hooft Commutation Relations}
\label{sub:hopf}

So far, the loop operators have been supported on the same circle. Let us now consider the composition of two supersymmetric loop operators supported on each of the two circles
\begin{alignat*}{3}
(i) \quad  x_1 &= b \cos \varphi  \qquad\qquad  &  (ii) \quad x_3 & = b^{-1} \cos\varphi \\ 
 x_2 &= b \sin \varphi    & x_4 & = b^{-1} \sin\varphi \\
 x_0 &= x_3 = x_4 = 0  \qquad\qquad & x_0 & = x_1 = x_2 = 0
\end{alignat*}
which are Hopf linked in the squashed three-sphere at the equator $\{x_0=0\}$. The loop operators supported on $(i)$ are those studied throughout this paper. They correspond to Verlinde operators constructed by transporting chiral primaries with momentum of the form $-b\lambda$ where $\lambda$ is a fundamental weight. Those supported on $(ii)$ are obtained by replacing $b \to 1/b$ in the difference operators. They correspond to transporting chiral primaries with momentum $ -\lambda / b$ around the same networks. 

In this subsection, let us denote the operators supported on the circles $(i)$ and $(ii)$ by $\cO^{(+)}$ and $\cO^{(-)}$ respectively. The operator $\cO^{(-)}$ is obtained from $\cO^{(+)}$ by replacing $b \to b^{-1}$. From the results above, it is now straightforward to compute the commutation properties of operators on Hopf linked circles. For all combinations of the loop operators constructed above we find
\be
\cO^{(+)}_{\lambda_e,\lambda_m} \, \cO^{(-)}_{\lb_e,\lb_m} = e^{2\pi i ( (\lb'_e,\lb_m) - (\lb_e,\lb'_m)  )} \, 
\cO^{(-)}_{\lb_e,\lb_m} \, \cO^{(+)}_{\lambda_e,\lambda_m} \, .
\label{'thooftcomm}
\ee 
Let us emphasize that these relations hold independent of the parameter $b$~\footnote{In reference~\cite{Drukker:2009id} it was demonstrated in the case of $g=A_1$ that the same relations hold for loop operators supported in the same circle in the limit $b=1$. The reason for this is that the difference operators for loop operators on circles $(i)$ and $(ii)$ coincide in this limit. We emphasize that for a pair of loop operators on Hopf linked circles, the commutation relations~\eqref{'thooftcomm} hold for any value of the parameter $b$.}. For the expectation value of two supersymmetric loop operators on Hopf linked circles to be well-defined, the corresponding difference operators should commute. This is the case provided that
\be
(\lb'_e,\lb_m) - (\lb_e,\lb'_m) \in \mathbb{Z} \, .
\label{mutuallocality}
\ee
This is how the mutual locality conditions are realised for supersymmetric loop operators on a four-sphere. 

Imposing the conditions~\eqref{mutuallocality}, the set of all loop operators given by pairs of weights $(\lb_e,\lb_e) \sim (w(\lb_e),w(\lb_m))$ with $w\in S_N$ can be partitioned into maximal sets of operators that are mutually local. Each maximal mutually local set defines a distinct realisation of the theory with gauge group of the form $SU(N) / H$ where $H\subset \mathbb{Z}_N$ and these sets form an intricate web under S-duality transformations~\cite{Gaiotto:2010be,Aharony:2013hda}. For theories of class $S$ without punctures, the origin of this structure has been understood from the perspective of the $(2,0)$ theory compactified on $C$ in reference~\cite{Tachikawa:2013hya} (see also the recent work~\cite{Xie:2013vfa}). It would be interesting to understand it better in the context of Liouville / Toda theory.

\section{Discussion}
\label{Section:Discussion}

The results of this paper have been summarised in the introduction~\ref{Section:Introduction}. Let us finish by pointing out what has not been accomplished in the present present paper and some suggestions for further study:

\begin{enumerate}
\item For topological defects / Verlinde operators corresponding to networks with junctions, we have been restricted to $A_2$ Toda theory. Further progress at higher rank requires the fusion and braiding matrices for four-point $\cW_N$-algebra conformal blocks involving completely degenerate momenta $\mu = -b\omega_j$ for all  anti-symmetric tensor representations $j=1,\ldots,N-1$.  In particular, one could then derive the rules for removing contractible networks and the generalised skein relations. We conjecture that they are given by the $A_{N-1}$-type spiders for the corresponding quantum group $U_{\q}(\mathrm{sl}(N))$ in the references~\cite{Kim:2006fk,Morrison:2007uq}. 
\item We did not find a convenient way to classify the irreducible networks on a given Riemann surface with punctures, even in the case of $A_{2}$ and a torus without puncture. This problem is much more involved than that of closed curves without intersections. Conjecturally, this classification should be the same as that of UV supersymmetric loop operators in theories of class $\cS$.
\item It would be interesting to study more examples in class $\cS$ theories with UV lagrangian descriptions. For example, with the present technology it is possible to study supersymmetric loop operators in $\cN=2$ linear quiver theories. This will require a better understanding of the exact computation of dyonic loop operators on a four-sphere in the cases where there is monopole bubbling.
\item The transformation properties of loop operators under $T:\tau \to \tau+1$ are straightforward to check by conjugating with the monodromy of the blocks $e^{2\pi i \Delta(\al)} \sim e^{i\pi (a,a)}$, corresponding to an $\cN=2$ supersymmetric Chern-Simons term at level 1 at the equator $\{x_0=0\}$. On the other hand, one must check that the partition function of the $T(\mathrm{SU}(N))$ theory on an S-duality domain wall intertwines the operators related by the transformation $S:\tau \to -1/\tau$. It would be interesting to check this property for the loop operators constructed here.
\item Our supersymmetric loop operators have been defined in the UV. It would be interesting understand how they decompose into IR loop operators under renromalisation group flow for $N>2$. For example, the three-sphere partition function of the renormalisation group domain wall for the $\cN=2^*$ theory should intertwine the UV and IR loop operators, as discussed for $N=2$ in references~\cite{Gaiotto:2010be,Dimofte:2011jd,Dimofte:2013lba}
\end{enumerate}

\acknowledgments

It is a pleasure to thank Fernando Alday, Martin Fluder, Jaume Gomis, Davide Gaiotto and Lotte Hollands for useful discussions. This work was partially completed at the Mathematical Institute, University of Oxford, where I was  supported by EPSRC grant EP/J019518/1. My research is supported by the Perimeter Institute for Theoretical Physics. Research at the Perimeter Institute is supported by the Government
of Canada through Industry Canada and by the Province of Ontario through the Ministry of Research and Innovation.


\appendix


\section{Group Theory Conventions}
\label{appendix:group}

We use the standard metric $(\, , \,)$ on the Cartan subalgebra of $A_{N-1}$ normalised such that the length $(e,e)=2$ for all roots $e$. The simple roots are denoted by $e_j$ where $j=1,\ldots,N-1$ and are dual to the fundamental weights $\omega_j$ where $j=1,\ldots,N-1$ in the sense that $(e_i,\omega_j)=\delta_{ij}$. The fundamental weights are the highest weights of the anti-symmetric tenor representations. The Weyl vector is given by the sum of the fundamental weights
\be
\rho = \sum_{j=1}^{N-1} \omega_j \, .
\ee
The weights of the fundamental representation are
\bea
h_j = \omega_1 - e_1 - \ldots - e_{j-1} \qquad j = 1,\ldots,N\, ,
\eea
and are normalised so that $(h_i,h_j)=\delta_{ij}-1/N$. The sum of the weights of the fundamental vanishes $\sum_{j=1}^N h_j=0$. The weights of the $r$-th anti-symmetric tensor representation are then given by $ h_{j_1}+\ldots+h_{j_r}$ for $1\geq j_1<\ldots<j_r \geq N$. The positive roots are given by $h_i-h_j$ for $i<j$ and the simple roots are $e_j=h_j-h_{j+1}$. The Weyl group $S_N$ acts by permutations of the weights of the fundamental representation.

When necessary, we choose a realisation of $A_{N-1}$ using traceless anti-hermitian matrices such that
\be
h_j = i \, \mathrm{diag} \Big( -\frac{1}{N}, \ldots, \underbrace{1-\frac{1}{N}}_j,\ldots,-\frac{1}{N} \, \Big) \, 
\ee
and define the metric by a trace $(a,b)=-\tr(ab)$ in the fundamental representation. For an element $a$ of the Cartan subalgebra we define its components by $a_j = (a,h_j)$ so that $\sum_{j=1}^N a_j = 0$ and $(a,a)=-\tr(a^2)=a_1^1+ \ldots + a_N^2$.


\section{Classical Toda and Spin Networks}
\label{Appendix:Classical}

In this appendix, we review the classical Toda equations on a Riemann surface $C$, their zero-curvature representation and the semi-classical analogue of the topological defect networks considered in the main text. 

Let us introduce complex coordinates $(z,\bar z)$ in a coordinate patch with flat background metric $g=dz \otimes d\bar z$. We consider a classical scalar field $\varphi(z,\bar{z})$ valued in the Cartan subalgebra of $A_{N-1}$ and obeys the Toda equations
\be
\del \bar\del \varphi = \sum_{j=1}^{N-1} e_j \exp{( e_j , \varphi ) } \, .
\label{eq:classicaltoda}
\ee
These equations are invariant under the conformal transformation $z \to w(z)$ with the scalar field transforming
\be
\varphi(z,\bar z) \to \varphi(z,\bar z) + \rho \, \log \left| \frac{\del w}{\del z} \right|^2\, .
\ee
In the case of $A_1$, there is only one component field $\varphi(z,\bar z)$ and we recover the classical Liouville equation.

The Toda equations imply the existence of a set of holomorphic conserved currents $w_j(z)$ of spins $j=2,\ldots,N$, which are invariant polynomials in the scalar field and its holomorphic derivatives $\varphi,\del\varphi,\ldots,\del^j\varphi$. These currents are neatly packaged in the generating operator
\bea
\prod_{j=0}^{N-1} (\del + (h_{N-j},\del \varphi) )  = \del^{N} + w_2 (z) \, \del^{N-2} + \ldots + w_N ( z) 
\eea
where $h_j$ are the weights of the fundamental representation. In particular, the holomorphic current
\be
w_2(z)  = ( \rho , \del^2 \varphi ) - \frac{1}{2} ( \del \varphi , \del \varphi )
\ee
is the holomorphic component of the stress-tensor and transforms as a projective connection under conformal transformations with central charge $c =N(N-1)(N+1)$. The remaining holomorphic currents $w_j(z)$ transform as tensors of spin $j=3,\ldots,N$. Finally, there are also anti-holomorphic currents $\overline{w}_j(\bar z)$ defined by an identical construction with the replacement $\del \to \bar\del$.

It is a difficult task to find solutions of the Toda equations. A useful strategy is to reformulate the problem in terms of  a flatness condition for an $SL(N)$ connection. The starting point is that the exponential $e^{-(\omega_1,\varphi)}$ obeys both holomorphic and anti-holomorphic differential equations
\bea
\left( \del^N + w_2 \, \del^{N-2} + \ldots + w_N \right) e^{-(\omega_1,\varphi)} &= 0 \, , \\
\left( \delbar^N + \tilde{w}_2 \, \delbar^{N-2} + \ldots + \tilde{w}_N \right)  e^{-(\omega_1,\varphi)}
 &= 0 \, .
 \label{opers}
\eea
as a consequence of the Toda equations~\eqref{eq:classicaltoda}. Thus, the exponential $e^{-(\omega_1,\varphi)}$ can be expressed in terms of linearly independent solutions $u_j(z)$ and $\tilde{u}_j(\bar z)$ to the differential equations~\eqref{opers},
\be
e^{-(\omega_1,\varphi)} = \sum_{j=1}^N u_j(z) \tilde{u}_j(\bar z)\, .
\ee
It can be shown that the remaining exponentials are given by
\be
e^{-(\omega_k,\varphi)} = \sum_{j_1 < \ldots < j_k}  
\begin{vmatrix}
u_{j_1} & \cdots & u_{j_k} \\
\del u_{j_1} & \cdots & \del u_{j_k} \\
\vdots &  & \vdots \\
\, \del^{k-1}u_{j_1} & \cdots & \del^{k-1}u_{j_k}  
\end{vmatrix} 
\begin{vmatrix}
\tilde{u}_{j_1} & \cdots & \tilde{u}_{j_k} \\
\delbar \tilde{u}_{j_1} & \cdots & \delbar \tilde{u}_{j_k} \\
\vdots &  & \vdots \\
\, \delbar^{k-1}\tilde{u}_{j_1} & \cdots & \delbar^{k-1}\tilde{u}_{j_k}  
\end{vmatrix} 
\label{sol}
\ee
together with the consistency condition that the Wronskian determinants obey the condition $W(u_1,\ldots, u_N)W(\tilde u_1 , \ldots, \tilde u_n) = 1$. The differential equations have monodromy and it is imperative that the solutions $u_j$ and $\tilde u_j$ have conjugate monodromy so that the solution for $\varphi(z,\bar z)$ is well-defined. In fact, this is used to determine the $w_j(z)$'s together with the boundary condition - see for example~\cite{Fateev:2007ab}.

The above construction can be reformulated in terms of a pair of flat connections. Given a solution $\varphi(z,\bar z)$ of the Toda equations with holomorphic currents $w_j(z)$ and anti-holomorphic currents $\bar{w}_j(\bar z)$, one can define a pair of $SL(N)$ flat connections given by
\be
A_z = 
\begin{pmatrix}
  0 & -1 &\cdots & 0 & 0 \\
  \vdots &  \ddots & \ddots & \ddots & \vdots \\
  0 & \cdots  & 0 & -1 & 0 \\
  0 & \cdots & 0 & 0 & -1 \\
  \,w_N \, & \cdots & \, w_3 \, & \, w_2 \, & 0
 \end{pmatrix} 
 \qquad A_{\bar z} = 0\, .
 \label{eq:fc1}
\ee  
and
\be
\tilde{A}_z = 0 \qquad \tilde{A}_{\bar z} = 
\begin{pmatrix}
  0 & -1 &\cdots & 0 & 0 \\
  \vdots &  \ddots & \ddots & \ddots & \vdots \\
  0 & \cdots  & 0 & -1 & 0 \\
  0 & \cdots & 0 & 0 & -1 \\
  \,\bar w_N \, & \cdots & \, \bar w_3 \, & \, \bar w_2 \, & 0
 \end{pmatrix} \, .
 \label{eq:fc2}
\ee  
Under changes of coordinate patch, the above connections transform by lower triangular matrices. The sections of the flat connections~\eqref{eq:fc1} and~\eqref{eq:fc2} are respectively of the form $s = (u,\del u, \ldots, \del^{N-1}u)$ and $\tilde{s} = (\tilde{u} , \delbar \tilde{u} ,\ldots, \delbar^N\tilde{u})$ where $u(z)$ and $\tilde{u}(\bar z)$ are solutions to the holomorphic differential equations~\eqref{opers}.

Let us introduce a basis of linearly independent sections $s_j(z)$ with $j=1,\ldots N$. Similarly, we introduce a basis of conjugate solutions $\tilde s_j(z)$. The Wronskians are then constructed using the invariant tensor of $SL(N,\mathbb{R})$
\be
W(u_1,\ldots,u_N) = ( s_1, \ldots s_N ) = \epsilon_{a_1\ldots a_N}  s_1^{a_1}(z) \ldots s_{N}^{a_N} (z)\, .
\ee
It is now clear that the solution~\eqref{sol} is expressed in terms of the sections $s_{j_1} \wedge \ldots \wedge s_{j_r}$ and $\tilde{s}_{j_1} \wedge \ldots \tilde{s}_{j_r}$ with $j_1 < \ldots < j_r$
of the pair of flat connections in the $r$-th anti-symmetric tensor representations.

The solution can be characterised by the monodromies of the pair of flat connections. Again, it is important that they have conjugate monodromy so that the solution for $\varphi(z,\bar z)$ is well-defined. In particular, the boundary condition at a puncture of $C$ is specified by fixing the trace of the holonomy in all of the anti-symmetric tensor representations,
\be
\tr_{\Lambda_r} \, P\, \exp  \oint_{C} A = \sum_{j_1 < \ldots < j_r} e^{-2 \pi (m_{j_1} + \cdots + m_{j_r}) }
\ee
for $r=1,\ldots,N-1$. This is a gauge invariant way to specify the holonomy eigenvalues $\{m_1,\ldots,m_N\}$ obeying $\sum_j m_j=0$ at each puncture.

Given a set of boundary conditions, the trace of the holonomy in the representations $\Lambda_r$ around all remaining homotopy classes of simple closed curves in $C$ is not sufficient to characterise the solution completely. One option is to consider closed curves with intersections. Another, that is perhaps more natural, is to specify the value of  `spin network' functionals constructed from the flat connection. Let us denote by
\be
U_{\Lambda_j}(z_1,z_2) = P \exp \int_{z_1}^{z_2} A
\ee
the parallel transport of a section in the fundamental representation $\Lambda_j$ from $z_1$ to $z_2$. One way to form a gauge invariant quantity is to let $z_2 \to z_1$ around a non-contractible closed curve $C$. Another is to contract three parallel propagators at a point using an invariant tensor on $\Lambda_i \times \Lambda_j \times \Lambda_k$. Such an invariant tensor exists if $i+j+k=N$. For example, in $A_2$ we can contract three in the fundamental at a point $z$ by
\be
\epsilon_{a_1 a_2 a_3} \, \epsilon^{b_1b_2b_3} \, {U(w_1,z)^{a_1}}_{b_1} \,{U(w_2,z)^{a_2}}_{b_2} \, {U(w_3,z)^{a_3}}_{b_3} \, .
\ee
Then, allowing all line to end on such vertices we obtain a spin network functional. Since the exponentials $e^{-(\omega_j,\varphi)}$ are the classical analogues of degenerate vertex operators labelled by the $j$-th anti-symmetric tensor representations, the spin network functionals can be see as the classical analogue of the Verlinde operators constructed in the main text.

The spin network functionals obey some basic properties that follow from the fact that they are constructed from matrices in $SL(N)$. Firstly $U_{\Lambda_j}(z_2,z_1) = U_{\Lambda_j}(z_1,z_2)^{-1}$ is linearly related to $U_{\Lambda_{N-j}}(z_1,z_2)$ using the invariant tensor $\epsilon^{a_1\ldots a_N}$. Thus reversing the orientation of a curve is equivalent to exchanging $j \leftrightarrow N-j$. Secondly, there are local relations due to properties of $\epsilon^{a_1\ldots a_N}$ and $\delta^a_b$. For example, for $A_1$ we have
\bea
\mathfig{0.08}{defects/contractible_loop_A1} = 2 \qquad\qquad  \mathfig{0.08}{defects/skein_abelian_2} = \, \mathfig{0.08}{defects/skein_abelian_3} + \,  \mathfig{0.08}{defects/skein_abelian_4}
\label{eq:classicalskein1}
\eea
as a consequence of the $SL(2)$ relations $\delta^a_a = 2$ and $\epsilon^{ab}\epsilon^{cd} + \epsilon^{bc} \epsilon^{ad} + \epsilon^{ca}\epsilon^{bd}= 0$ respectively. Similarly, for $A_2$ we have
\bea
& \mathfig{0.08}{defects/contractible_loop_A2} = \, 3 \qquad\quad \mathfig{0.15}{defects/contractible_bubble} = \, -2 \, \mathfig{0.13}{defects/line}
\\
& \mathfig{0.45}{defects/contractible_square}
\\
& \mathfig{0.09}{defects/skein_2} = \,\mathfig{0.09}{defects/skein_3} + \,   \mathfig{0.09}{defects/skein_4}
\label{eq:classicalskein2}
\eea
from identities involving $\epsilon^{abc}$ and $\delta^a_b$. Note that the ordering of the intersections is irrelevant and that any closed curves with intersections can be decomposed into networks. These relations are the classical analogues of the quantum skein relations found for the Verlinde operators in the main text.


\section{Generalised Hypergeometric Equation}
\label{appendix:hypergeometric}

The hypergeometric equation of order $N$
\be
\left[ z \prod_{j=1}^N\left( z\frac{d}{d z}+\sigma_j \right) -\prod_{j=1}^N \left( z\frac{d}{dz} -\tau_j \right) \right] y = 0 
\ee
depends on $2N$ parameters $\{\sigma_j,\tau_j\}$ has regular singularities at points $z=0,1,\infty$. There are $N$ linearly independent solutions defined by hypergeometric series around $z=0$
\be
\begin{aligned}
y_i & = \, z^{\tau_i} \, {}_{N-1}F_N\left( \begin{array}{cc} \tau_i +\sigma_1,\ldots,\tau_i+\sigma_N \\ \tau_i-\tau_1+1,\ldots,\tau_i-\tau_N+1  \end{array} |z\right) \\
& = \, z^{\tau_i} \sum_{\nu=0}^{\infty} \prod_{j=1}^N \frac{(\tau_i-\sigma_j)_{\nu}}{(\tau_i-\tau_j+1)_{\nu}} z^{\nu}
\end{aligned}
\label{ys}
\ee
where $(\al)_{\nu}=\Gamma(\al+\nu) / \Gamma(\al)$ is the Pochhammer symbol. The hypergeometric series are convergent for $|z|<1$ with generic parameters $\{\sigma_j,\tau_j\}$. Notice that the solutions are uniquely distinguished by their monodromy $M_j^{(0)}=e^{2 \pi i \tau_j}$ around the regular singular point $z=0$. 

Since the hypergeometric equation is invariant under the interchange of $z\leftrightarrow1/z$ and $\sigma_j \leftrightarrow \tau_j$ there is another basis of linearly independent solutions defined by hypergeometric series around $z=\infty$
\be
\begin{aligned}
\bar y_i & = \, z^{-\sigma_i} \, {}_{N-1}F_N\left( \begin{array}{cc} \sigma_i +\tau_1,\ldots,\sigma_i+\tau_N \\ \sigma_i-\sigma_1+1,\ldots,\sigma_i-\sigma_N+1  \end{array}|\frac{1}{z}\right) \\
& = \, z^{-\sigma_i} \sum_{\nu=0}^{\infty} \prod_{j=1}^N \frac{(\sigma_i-\tau_j)_{\nu}}{(\sigma_i-\sigma_j+1)_{\nu}} z^{-\nu}
\end{aligned}
\label{ybars}
\ee
which are generically convergent for $|z|>1$. The solutions are now uniquely characterised by their monodromy $M_j^{(\infty)}=e^{2 \pi i \sigma_j}$ around $z=\infty$.

The solutions \eqref{ys} and \eqref{ybars} have an analytic continuation using the Mellin-Barnes integral representation of the hypergeometric function to $z\in \mathbb{C}-[1,\infty)$ and $z\in\mathbb{C}-(0,1]$ respectively. In order to relate the solutions on the overlap $z\notin\mathbb{R}^+$, it is convenient to introduce the normalised solutions
\be
Y_i = \prod_{k=1}^N \frac{\Gamma(\tau_i+\sigma_k)}{\Gamma(\tau_i-\tau_k+1)} \, y_i \qquad
\bar{Y}_i = \prod_{k=1}^N \frac{\Gamma(\sigma_i+\tau_k)}{\Gamma(\sigma_i-\sigma_k+1)} \, \bar{y}_i \, .
\label{normsol}
\ee
The transformation between them is
\be
\begin{aligned}
Y_i  = \sum_{j=1}^N \frac{ e^{ i \pi \epsilon (\tau_i+\sigma_j)}}{\sin\pi(\tau_i+\sigma_j)} \,  \frac{ \prod\limits_{k=1}^{N} \sin\pi(\sigma_j+\tau_k)}{\prod\limits_{k\neq j}^N \sin\pi(\sigma_k-\sigma_j)} \, \bar{Y}_j \\
\bar{Y}_i  = \sum_{j=1}^N \frac{ e^{ - i \pi \epsilon (\sigma_i+\tau_j)}}{\sin\pi(\sigma_i+\tau_j)} \,  \frac{ \prod\limits_{k=1}^{N} \sin\pi(\tau_j+\sigma_k)}{\prod\limits_{k \neq j}^N \sin\pi(\tau_k-\tau_j)} \, Y_j
\end{aligned}
\label{sufusion}
\ee
where $\epsilon=\mbox{Sgn} \, \mbox{Im} (z)$. Note that the inverse transformation is obtained by interchanging $z\leftrightarrow1/z$ and $\sigma_j \leftrightarrow \tau_j$. 

The solutions with definite monodromy around $z=1$, however, do not have representations as hypergeometric series in the variable $(1-z)$, except in the case $N=2$. In any case, the monodromy matrix around $z=1$ has determinant
\be
\det M_{(1)} = e^{-2 \pi i \Sigma}\, .
\ee
where $\Sigma=\sum_{j=1}^n (\sigma_j+\tau_j)$. There is one solution with monodromy $e^{-2 \pi i \Sigma}$ and an $(n-1)$ dimensional subspace with degenerate monodromy $1$. The eigenfunctions most simply expressed as linear combinations of the normalised solutions~\eqref{normsol}. The solution with monodromy $e^{-2 \pi i \Sigma}$ around $z=1$ is given by the combination
\be
\begin{aligned}
X_1 & = \sum_{j=1}^N \,  \frac{ \prod\limits_{k=1}^N \sin\pi(\sigma_k+\tau_j)}{ \prod\limits_{k\neq j}^N\sin\pi(\tau_k-\tau_j)} \, Y_j \\
\end{aligned}
\ee
and the degenerate subspace with monodromy $1$ around $z=1$ is spanned by the combinations
\be
\begin{aligned}
X_i = Y_1 - Y_i \qquad i=2,\ldots,N \, .
\end{aligned}
\ee
The determinant of the transformation matrix is $\sin(\pi\Sigma)$ and the inverse transformations are
\bea
Y_1 & = \frac{1}{\sin \pi (N-\Sigma)} \Bigg[ X_1 + \sum_{j=2}^N \, \frac{ \prod\limits_{k=1}^N \sin\pi(\sigma_k+\tau_j)}{ \prod\limits_{k\neq j}^N\sin\pi(\tau_k-\tau_j)}  \, X_j \Bigg] \\
Y_i & = \frac{1}{\sin \pi(N-\Sigma)} \Bigg[ X_1 + \sum_{j\neq 1,i}^N \frac{ \prod\limits_{k=1}^N\sin\pi(\sigma_k+\tau_j)}{ \prod\limits_{k\neq j}^N\sin\pi(\tau_k-\tau_j)} X_j  - \sum_{j\neq i}^N \frac{ \prod\limits_{k=1}^N\sin\pi(\sigma_k+\tau_j)}{ \prod\limits_{k\neq j}^N\sin\pi(\tau_k-\tau_j)} X_i  \Bigg] 
\label{stinverse}
\eea
where we have absorbed a sign by writing $(-1)^{N-1}\sin\pi \Sigma=\sin\pi(N-\Sigma)$.


\section{Fusion and Braiding}
\label{appendix:fusion}

The fusion and braiding matrices needed in this paper can be derived from a four-point correlation function involving one completely degenerate vertex operator $V_{\mu}$ with momentum $\mu=-b\omega_1$, one semi-degenerate operator $V_{\nu}$ with momentum $\nu=\kappa\, \omega_{N\!-\!1}$, and two non-degenerate vertex operators with momentum $\al_1$ and $2Q-\al_2$ - see figure. 

\textit{Figure}

The authors~\cite{Gomis:2010kv} of have shown that such a correlation function must take the form
\be
\la V_{\bar\al_2}(\infty)V_{\nu}(1)V_{\mu}(z) V_{\al_1}(0) \ra = |z|^{-2 b (\mu-Q,h_1)}|1-z|^{2b(\nu,h_1)} G(z,\bar{z})
\label{4point}
\ee 
where the function $G(z,\bar{z})$ is constructed from diagonal combinations of solutions to the generalised hypergeometric differential equation
\be
\left[ z \prod_{j=1}^N\left( z\frac{d}{d z}+\sigma_j \right) -\prod_{j=1}^N\left( z\frac{d}{dz} -\tau_j \right) \right] G(z,\bar z) = 0 
\ee
and its conjugate with $z\leftrightarrow\bar{z}$ but the same parameters $\{\sigma_j,\tau_j\}$. The parameters of the hypergeometric equation are related to the Toda momenta by
\be
\begin{aligned}
\sigma_j & = b(\bar\al_2-Q,h_j) + b (\nu,h_1) \\
& = - i b \, a_{2,j} +\frac{b\,\kappa}{N}\end{aligned}
\ee
and
\be
\begin{aligned}
\tau_j &= b(\al_1-Q,h_j) + b (\mu,h_1) \\
& = i b\, a_{1,j}- b^2 \left( 1-\frac{1}{N} \right)
\end{aligned}
\ee
where for non-degenerate momenta $\al = Q + i a$ and $a_j = (a,h_j)$. The hypergeometric equation has regular singularities at points $z=0,1,\infty$ corresponding to the three operator product expansion channels. The corresponding $W_N$-algebra conformal blocks in each channel can be constructed from the solutions in appendix~\ref{appendix:hypergeometric}. The linear transformations between the bases provide the fusion and braiding matrices for the corresponding $W_N$-algebra conformal blocks.

The solutions to the hypergeometric equation with diagonal monodromy around $z=0$ and $z=\infty$ are hypergeometric functions with argument $z$ and $1/z$ respectively. The corresponding conformal blocks 
\be
\begin{aligned}
\mathfig{0.25}{blocks/schannel_generic_j} \hspace{-8mm} & = z^{s_j} \, (1-z)^{b\kappa/N} \, {}_{N-1}F_N \left( \begin{array}{cc} \tau_j +\sigma_1,\ldots,\tau_j+\sigma_N \\ \tau_j-\tau_1+1,\ldots,\tau_j-\tau_N+1  \end{array} |\, z\right) \\
\mathfig{0.25}{blocks/uchannel_generic_j} \hspace{-8mm}  & = z^{-u_j}\left(\frac{z-1}{z} \right)^{b\kappa/N} {}_{N-1}F_N \left( \begin{array}{cc} \sigma_j +\tau_1,\ldots,\sigma_j+\tau_N \\ \sigma_j-\sigma_1+1,\ldots,\sigma_j-\sigma_N+1  \end{array}|\, \frac{1}{z}\right)
\end{aligned}
\ee
appear in the expansion of the 4-point correlation function~\eqref{4point} using the operator product expansions
\be
\begin{aligned}
V_{\mu}(z) \cdot V_{\al_1}(0) & = \sum_{j=1}^n C_{\mu,\al_1}^{\al_1-b h_j} \left( V_{\al_1-b h_j} |z|^{2 s_j} + \ldots \right) \\
V_{\mu}(z) \cdot V_{\bar\al_2}(\infty) & = \sum_{j=1}^n C_{\mu,\bar\al_2}^{\bar\al_2-b h_j}\left( V_{\bar\al_2+b h_j} |z|^{-2 u_j} + \ldots \right)
\end{aligned}
\ee
where the leading powers are
\be
\begin{aligned}
s_j & = \Delta(\al_1-bh_j)-\Delta(\al_1)-\Delta(\mu) \\
& = i b a_{1,j} + \frac{1}{2}bq(N-1) \\
u_j  & =\Delta(\al_2+bh_j)-\Delta(\al_2)+\Delta(\mu) \\
& = -i a_{2,j}-\frac{1}{2}bq(N-1)-b^2\left(1-\frac{1}{N} \right)\, .
\end{aligned}
\ee
The s- and u-channel conformal blocks are uniquely identified by their monodromies $M^{(0)}_j = e^{2 \pi i s_j}$ and $M^{(\infty)}_j=e^{2 \pi i u_j }$ around the regular singular points $z=0$ and $z=\infty$ respectively. 

The above conformal blocks can have analytic continuations via the Mellin-Barnes integral representation and in the region $z\notin \mathbb{R}^+$ there is a linear transformation between the two bases of solutions. From appendix~\ref{appendix:hypergeometric} we find
\bea
\mathfig{0.25}{blocks/schannel_generic_i} \hspace{-8mm}  &=  \sum_{j=1}^n B^{(\epsilon)}_{ij} \mathfig{0.25}{blocks/uchannel_generic_j} \hspace{-8mm} \\
\mathfig{0.25}{blocks/uchannel_generic_i} \hspace{-8mm} & =  \sum_{j=1}^n B^{(\epsilon)-1}_{ij} \mathfig{0.25}{blocks/schannel_generic_j} \hspace{-8mm} 
\eea
where
\bea
B^{(\epsilon)}_{ij} & = e^{-i \pi \epsilon b  \kappa/N}\prod_{k=1}^N \frac{\Gamma(\tau_i-\tau_k+1)}{\Gamma(\tau_i+\sigma_k)} \cdot \frac{\pi e^{i \pi \epsilon(\tau_i+\sigma_j)}}{\sin\pi(\sigma_i+\tau_j)} \cdot \prod_{k=1}^N \frac{\Gamma(\sigma_k-\sigma_j)}{\Gamma(1-\tau_k-\sigma_j)} \\
& = e^{i \pi \epsilon (\tau_i+\sigma_j - b\kappa / N )} \prod_{k\neq i}^N \, \frac{\Gamma(1-\tau_k+\tau_i)}{\Gamma(1-\tau_k-\sigma_j)}  \, \prod_{k\neq j}^N \,  \frac{\Gamma(\sigma_k-\sigma_j)}{\Gamma(\sigma_k+\tau_i)}
\eea
and
\bea
B^{(\epsilon)-1}_{ij} & = e^{i \pi \epsilon b \kappa/n}\prod_{k=1}^N \frac{\Gamma(\sigma_i-\sigma_k+1)}{\Gamma(\sigma_i+\tau_k)} \cdot \frac{\pi e^{-i \pi \epsilon(\sigma_i+\tau_j)}}{\sin\pi(\tau_i+\sigma_j)} \cdot \prod_{k=1}^N \frac{\Gamma(\tau_k-\tau_j)}{\Gamma(1-\sigma_k-\tau_j)} \\
& = e^{-i \pi \epsilon (\tau_i+\sigma_j - b \kappa/N)} \, \prod_{k\neq i}^N \, \frac{\Gamma(1-\sigma_k+\sigma_i)}{\Gamma(1-\sigma_k-\tau_j)} \, \prod_{k\neq j}^N \, \frac{\Gamma(\tau_k-\tau_j)}{\Gamma(\tau_k+\sigma_i)}
\eea
and $\epsilon$ is the sign of the imaginary part of $z$. Note that the inverse transformation matrix $B_{ij}^{(\epsilon)-1}$ is obtained by exchanging $\sigma_j\leftrightarrow\tau_j$ and $z\leftrightarrow1/z$. 

The t-channel conformal blocks cannot be expressed as generalised hypergeometric functions with argument $(1-z)$, except in the special case $n=2$. However, they can be constructed by forming linear combinations of the s-channel conformal blocks that are eigenfunctions of the monodromy operator around $z=1$. The monodromy matrix around $z=1$ has determinant 
\be
\det\, M^{(1)} = e^{2 \pi i (b \kappa - \Sigma) }
\ee
where 
\be
\Sigma = \sum_{j}(\sigma_j+\tau_j) = - b^2(n-1) + b \kappa
\ee
It has one non-degenerate eigenvalue $e^{2 \pi i b ( \kappa /N - \Sigma )}$ and $(N-1)$ degenerate eigenvalues $e^{2 \pi i b \kappa/N}$. On the other hand, the operator product expansion predicts the appearance of two primary operators
\be
\begin{aligned}
V_{\mu} \cdot V_{\nu} & = \, C_{\mu,\nu}^{\nu-b h_1} \left( \, |1-z|^{2 t_1} \, V_{\nu-bh_1} + \cdots \right) \\
& + \,  C_{\mu,\nu}^{\nu-b h_N} \left( \, |1-z|^{2 t_N} \, V_{\nu-b h_N} + \cdots \right)
\end{aligned}
\ee
where
\be
\begin{aligned}
t_1 & = b \kappa/N \\
t_2 & = q b(N-1) - b\kappa(1-1/N)\, .
\end{aligned}
\ee
Therefore the monodromies predicted by the operator product expansion, $e^{2 \pi i t_1} = e^{2 \pi i b \kappa/N}$ and $e^{2 \pi i t_2} = e^{2\pi i b (\kappa/N - \Sigma)}$ are in agreement with the eigenvalues of the monodromy operator. 

We now introduce a basis of conformal blocks that are eigenfunctions of the monodromy operator around $z=1$. The unique eigenfunction with monodromy $e^{2 \pi i (b \kappa/N-\Sigma)}$ is
\be
\mathfig{0.23}{blocks/tchannel_generic_1} \hspace{-6mm} = \Gamma(n-\Sigma) \sum_{j=1}^N  \prod_{k=1}^N  \frac{\Gamma(\tau_k-\tau_j)}{\Gamma(1-\sigma_k-\tau_j)} \; \mathfig{0.25}{blocks/schannel_generic_j} \hspace{-8mm}
\ee
where the normalisation has been chosen such that the leading coefficient in a series expansion in $(1-z)$ is 1. It corresponds to the propagation of the primary operator $V_{\nu-bh_N}$ in the intermediate channel. A basis of solutions with monodromy $e^{2 \pi i b \kappa/N}$ is given by the following combinations of s-channel blocks
\be
 \Gamma(n-\Sigma) \Bigg[ \;\prod_{k=1}^N \frac{\Gamma(\tau_1+\sigma_k)}{\Gamma(\tau_1-\tau_k+1)} \mathfig{0.25}{blocks/schannel_generic_1} \hspace{-8mm}- \prod_{k=1}^N \frac{\Gamma(\tau_j+\sigma_k)}{\Gamma(\tau_j-\tau_k+1)} \mathfig{0.25}{blocks/schannel_generic_j} \hspace{-4mm} \Bigg]
\ee
for $j=2,\ldots,N$. The inverse transformations for this particular basis can be determined from equation~\eqref{stinverse} in appendix~\ref{appendix:hypergeometric}. The required combination of t-conformal blocks in the subspace of monodromy $e^{2\pi i b \kappa/N}$ depends on $\kappa$ and the external momenta $\al_1$ and $\al_2$. In the example studied in the main text, the required t-channel block is uniquely determined by the allowed momenta in the s-channel.



\section{Factorizations of the Toda 3-pt Function}
\label{appendix:special}

\textit{The results summarised in this appendix were obtained in collaboration with Martin Fluder, Lotte Hollands and Paul Richmond in the course of a forthcoming paper on surface defects and the superconformal index~\cite{sd} where further explanations and derivations can be found.}
\medskip

Let us briefly review some properties of special functions. We fix $b\in \mathbb{R}_{>0}$ and define $q\equiv b+b^{-1}$ as in the main text. The double gamma function $\Gamma_b(x)$ is a meromorphic function of $x$ characterised by the functional equation
\be
\Gamma_b(x+b) = \sqrt{2\pi}\, b^{bx-\frac{1}{2}}  
\Gamma_b(x) / \Gamma(bx)
\ee
where $\Gamma(x)$ is the Euler gamma function and its value $\Gamma_b(q/2)=1$. The double sine function is then a meromorphic function defined by the formula $S_b(x) \equiv \Gamma_b(x) / \Gamma_b(q-x)$ and is characterised by the functional equation
\be
S_b(x+b)=2\sin(\pi b x)S_b(x)\, .
\ee
We will also need the function $\Upsilon_b(x)^{-1}=\Gamma_b(x)\Gamma_b(q-x)$ which is entire analytic.

Consider the three-point function $C(\al,2Q-\al,\nu)$ in $A_{N-1}$ Toda theory where two legs have opposite momentum forming an internal channel. The momentum in the internal channel $\al=Q+ia$ with $a\in \mathbb{R}$ describes non-degenerate and delta-function normalisable states and the momentum $\nu=N(q/2+im)\omega_{N-1}$ with $m\in \mathbb{R}$ is semi-degenerate. Substituting these momenta into the more general result of~\cite{Fateev:2005gs,Fateev:2007ab} we find that
\bea
C(\al,2Q-\al,\nu) & = f(m) \frac{ \prod\limits_{i<j}^N \Upsilon_b\left( ia_{ij} \right)\Upsilon_b\left( -ia_{ij} \right)}{\prod\limits_{i,j=1}^N \Upsilon_b\left( \frac{q}{2} + ia_{ij} + im \right)} 
\eea
where $a_{ij} = a_i-a_j$ and the proportionality factor $f(m)$ is independent of the internal parameter $a$. Since we are concerned with difference operators acting on $a$ this factor is not important here.

We now consider two different ways to factorise the three-point function and absorb it into the $W_N$-algebra conformal blocks. The first maximally simplifies the Verlinde operators and coincides with a common normalisation of Virasoro conformal blocks in Liouville theory in the case $N=2$. It is expected that this corresponds to a half-sphere partition function of $\cN=2^*$ theory with Dirichlet boundary conditions for the vectormultiplet at the equator. The second normalisation corresponds to computing the Nekrasov partition function of the $\cN=2^*$ theory with deformation parameters $\epsilon_1 = b$ and $\epsilon_2=b^{-1}$.

\subsection*{Renormalised Conformal Blocks}

Let us now express the Toda three-point function in terms of double gamma functions and manipulate the answer into a convenient form using the functional equation. We find that it can be expressed as
\bea
\mu(a) \left| \, \frac{\prod\limits_{i,j=1}^N \Gamma_b\left( \frac{q}{2} +ia_{ij}+im \right) }{\prod\limits_{i<j}^N\Gamma_b\left( q + ia_{ij} \right) \Gamma_b(q-ia_{ij} )}\, \right|^2 \, .
\eea
where
\be
\mu(a) = \prod_{i<j} \, 2\sinh\left( \pi ba_{ij} \right) \,  2\sinh\left( \pi b^{-1} a_{ij} \right)
\ee
is the partition function of an $\cN=2$ vectormultiplet on a squashed three-sphere~\cite{Hama:2011ea}, which is identified here with the equator $\{x_0=0\}$.

As described in the main text, we can now absorb the three-point function into the $W_N$-algebra conformal blocks, by defining new renormalised blocks
\be
\cG_{a,m}(\tau) = \frac{ \prod\limits_{i,j=1}^N \Gamma_b\left( \frac{q}{2} +ia_{ij}+im \right)  }{ \prod\limits_{i<j}^N \Gamma_b\left( q + ia_{ij} \right) \Gamma_b(q-ia_{ij} ) } \, \cF_{\al,\nu}(\tau) \, .
\label{FtoG}
\ee
so that the full correlation function is
\be
\int da \, \mu(a) | \, \cG_{a,m}(\tau) \, |^2 \,.
\ee
Thus, we believe that the renormalised conformal block corresponds to the partition function on $\{x_0>0\}$ with Dirichlet boundary conditions for the vectormultiplet at the equator. Thus, in order to transform between Verlinde operators acting on $\cF_{\al,\nu}(\tau)$ and those acting on $\cG_{a,m}(\tau)$ we have to conjugate by the factor in equation~\eqref{FtoG}. 

Let us concentrate on the Verlinde operator corresponding to the fundamental 't Hooft loop. Acting on the unnormalised conformal blocks $\cF_{\al,\nu}(\tau)$, this Verlinde operator has been computed in~\cite{Gomis:2010kv} and the result is given by
\be
\sum_{j=1}^N \left[ \, \prod_{k\neq j}^N \frac{\Gamma\left(i ba_{kj}\right)}{\Gamma\left( \frac{bq}{2}+iba_{kj}-i b m \right)} \frac{\Gamma\left(b q+i ba_{kj} \right)}{\Gamma\left( \frac{bq}{2}+iba_{kj}+i b m \right)} \,  \right] \Delta_j
\label{unnorm}
\ee
where
\be
\Delta_j:a_j\to a_i+ib(\delta_{ij}-1/N) \, .
\ee
We now conjugate this operator by the one-loop factor normalisation factor in equation~\eqref{FtoG}. By repeated application of the functional equation for the double gamma function, we find the result
\begin{multline}
\sum_{j=1}^N \left[ \, \prod_{k\neq j}^N \frac{\Gamma\left(i ba_{kj}\right)\Gamma\left( 1-iba_{kj} \right)}{\Gamma\left( \frac{bq}{2}+iba_{kj}-i b m \right)\Gamma\left( 1-\frac{qb}{2}-iba_{kj} +ibm\right)} \right]  \Delta_j \\
 = \, \sum_{j=1}^N \left[ \,  \prod_{k\neq j}^N \frac{\sin\pi b\left( \frac{q}{2} + ia_{kj}-im \right)}{\sin \pi b \left( i a_{kj}  \right)}  \right] \Delta_j
 \label{Gop}
\end{multline}
as claimed in the main text. With patient bookkeeping, the same computation can be performed for the difference operators in completely anti-symmetric tensor representations.

\subsection*{Nekrasov Partition Function}

For comparison with the exact computation of 't Hooft loop on the four-sphere in~\cite{Gomis:2011pf}, it is necessary to consider another factorisation of the Toda three-point function, such that the difference operators act on the Nekrasov partition function with $\epsilon_1 = b$ and $\epsilon_2 = b^{-1}$, which we denote by $\cZ_{a,m}(\tau)$. 

Thus we now express the three-point function
\be
C(\al,2Q-\al,\nu) = f(m)\,  |\, \cZ^{1-\mathrm{loop}}_{a,m}(\tau) \, |^2 
\ee
where 
\be
\cZ^{1-\mathrm{loop}}_{a,m}(\tau) = \left[ \, \frac{ \prod\limits_{i<j}^N \Upsilon_b\left( ia_{ij} \right) \Upsilon_b\left( -ia_{ij} \right) }{ \prod\limits_{i,j=1}^N \Upsilon_b\left( \frac{q}{2}+ia_{ij}+im \right)} \, \right]^{1/2}
\ee
are the one-loop contributions to the Nekrasov partition function. The classical and instanton contributions to the Nekrasov partition function are encoded in the $W_N$-algebra conformal blocks. Thus, modulo the factor $f(m)$, the complete Toda correlator can be expressed
\be
\int da | \, \cZ_{a,m}(\tau) \, |^2
\ee
in agreement with the exact computation of the partition function of the $\cN=2^*$ theory on an ellipsoid in~\cite{Hama:2012bg}.

To obtain difference operators acting on the Nekrasov partition function, it is easier at this stage to start from the relationship to the renormalised $W_N$-algebra conformal blocks. In fact, from the relationship between the double gamma double sine functions and upsilon functions, we find
\be
\cZ_{a,m}(\tau) = \left[ \, \frac{\prod\limits_{i<j}^N S_b(q+ia_{ij}) S_b(q-ia_{ij} ) } { \prod\limits_{i,j=1}^N S_b(\frac{q}{2}+ia_{ij}+im )} \, \right]^{1/2} \cG_{a,m}(\tau) \, .
\label{ZtoG}
\ee
Again, the difference operator that acts on the Nekrasov partition function is obtained by conjugation by this factor. By repeated use of the functional equation for the double sine function, we now find the operator
\be
\sum_{j=1}^k \left[ \; \prod_{k \neq j}^N \frac{\sin\pi b \left( \frac{q}{2} +ia_{kj}+im \right) \sin\pi b\left( \frac{q}{2} + ia_{kj} -im\right)}{ \sin\pi b\left(ia_{kj}\right) \sin\pi b( q+ia_{kj} )}  \, \right]^{1/2}  \Delta_j \, .
\ee
which agrees with the exact computation of the fundamental 't Hooft loop operator in the case of a round four-sphere~\cite{Gomis:2011pf}. With patient bookkeeping, a similar result can be reached for the Verlinde operators corresponding to 't Hooft loops labelled by any anti-symmetric tensor representation.

\bibliographystyle{JHEP}
\bibliography{networkdefects}
\end{document}